\documentclass[prd,amsfonts,onecolumn,superscriptaddress,aps,nofootinbib,11pt]{revtex4}

\pdfoutput=1
\usepackage{amsmath,amssymb,mathrsfs}
\usepackage{graphicx}
\usepackage{enumerate}
\usepackage[usenames,dvipsnames]{xcolor} 
\usepackage{hyperref}
\hypersetup{colorlinks=true,
    breaklinks=true,
    linkcolor=Blue,          
    citecolor=Plum,        
    urlcolor=YellowOrange           
}
\usepackage[]{units}
\usepackage{ulem}

\providecommand{\eq}[1]{\begin{equation} #1 \end{equation}}
\providecommand{\eqali}[1]{\begin{equation}\begin{aligned} #1
    \end{aligned}\end{equation}}
\providecommand{\lag}{\mathcal{L}}

\providecommand{\ums}[2][1]{\ml{\tfrac{#1}{#2}}}

\providecommand{\ZZ}{\mathbb{Z}}
\providecommand{\ml}[1]{\mbox{\large $#1$}}
\providecommand{\mss}[1]{\mbox{\scriptsize $#1$}}
\providecommand{\aver}[1]{\langle #1 \rangle}
\providecommand{\tp}{{\mss{\mathsf{T}}}}
\providecommand{\mtrx}[1]{\begin{pmatrix} #1 \end{pmatrix}}
\providecommand{\cG}{\mathcal{G}}
\providecommand{\cE}{\mathcal{E}}
\providecommand{\cU}{\mathcal{U}}
\providecommand{\cD}{\mathcal{D}}
\providecommand{\cM}{\mathcal{M}}
\providecommand{\tM}{\tilde{M}}
\providecommand{\bh}{\bar{h}}

\DeclareMathOperator{\re}{\mathrm{Re}} 
\DeclareMathOperator{\im}{\mathrm{Im}}

\providecommand{\tphi}{\tilde{\phi}}
\providecommand{\eps}{\epsilon}

\providecommand{\to}{\rightarrow}

\DeclareMathOperator{\diag}{diag}

\usepackage{tabu}
\providecommand{\be}{ \begin{equation} }
\providecommand{\ee}{\end{equation}}
\providecommand{\bea}{\begin{eqnarray}}
\providecommand{\eea}{\end{eqnarray}}
\providecommand{\nn}{\nonumber}

\begin{document}
\title{Hierarchical Fermions and Detectable $Z^\prime$\\ from Effective Two-Higgs-Triplet 3-3-1 Model}

\author{E.~R.~Barreto}%
\email{elmerraba@gmail.com}
\affiliation{Centro de Ci\^encias Naturais e Humanas, Universidade Federal do ABC,\\ Santo Andr\'e 09210-580, S\~ao Paulo, Brasil}

\author{A.~G.~Dias}%
\email{alex.dias@ufabc.edu.br}
\affiliation{Centro de Ci\^encias Naturais e Humanas, Universidade Federal do ABC,\\ Santo Andr\'e 09210-580, S\~ao Paulo, Brasil}

\author{J.~Leite}%
\email{julio.leite@ufabc.edu.br}
\affiliation{Centro de Ci\^encias Naturais e Humanas, Universidade Federal do ABC,\\ Santo Andr\'e 09210-580, S\~ao Paulo, Brasil}

\author{C.~C.~Nishi}%
\email{celso.nishi@ufabc.edu.br}
\affiliation{Centro de Matem\'atica, Computa\c{c}\~ao e Cogni\c{c}\~ao, Universidade Federal do ABC,\\ Santo Andr\'e 09210-580, S\~ao Paulo, Brasil}

\author{R.~L.~N.~Oliveira}%
\email{robertol@ifi.unicamp.br} 
\affiliation{Centro de Ci\^encias Naturais e Humanas, Universidade Federal do ABC,\\ Santo Andr\'e 09210-580, S\~ao Paulo, Brasil}

\author{W.~C.~Vieira}%
\email{william.vieira19@gmail.com} 
\affiliation{Centro de Ci\^encias Naturais e Humanas, Universidade Federal do ABC,\\ Santo Andr\'e 09210-580, S\~ao Paulo, Brasil}

\date{\today}

\begin{abstract}

We develop a ${\rm  SU}(3)_C\otimes {\rm  SU}(3)_L\otimes {\rm
U}(1)_X$ model where the number of fermion
generations is fixed by cancellation of gauge anomalies, being a type
of 3-3-1 model with new charged leptons. Similarly to the economical
3-3-1 models, symmetry breaking is achieved effectively with two
scalar triplets so that the spectrum of scalar particles
at the TeV scale
contains just two CP even scalars, one of which is the
recently discovered Higgs boson, plus a charged scalar.
Such a scalar sector is simpler than the one in the Two Higgs Doublet
Model, hence more attractive for phenomenological studies, and has no
FCNC mediated by scalars except for the ones induced by the mixing of SM fermions with
heavy fermions.
We identify a global residual symmetry of the model which guarantees
mass degeneracies and some massless fermions whose masses need to be
generated by the introduction of effective operators. The fermion
masses so generated require less fine-tuning for most of the SM fermions and
FCNC are naturally suppressed by the small mixing between the third
family of quarks and the rest.
The effective setting is justified by an ultraviolet completion of the
model from which the effective operators emerge naturally. A detailed
particle mass spectrum is presented, and an analysis of the $Z^\prime$
production at the LHC run II is performed to show that it could be easily
detected by considering the invariant mass and transverse momentum distributions 
in the dimuon channel.

\end{abstract}
\maketitle
\section{Introduction}

The measurements of the Higgs boson properties and their actual agreement with the Standard Model predictions~\cite{Aad:2012tfa,Chatrchyan:2012xdj,Aad:2015zhl,Aaboud:2017xsd} have corroborated the simplest implementation of the Higgs mechanism as the source of electroweak symmetry breaking. Although the CERN Large Hadron Collider (LHC) has not yet provided clear evidence 
for new physics, the Standard Model (SM) consolidation has helped us to put in a firmer 
footing a series of its theoretical problems such as the severe hierarchy of 
the Yukawa couplings (the flavor problem), 
including the neutrino masses and mixing problem; the number of fermion generations; the chiral nature 
of the electroweak interaction;  matter-antimatter asymmetry of the Universe; the strong CP problem; the dark matter 
content of the Universe; and the vacuum stability. The seeking of solutions for one or more of these problems has often guided the 
development of new models by extending the field content of the SM or, sometimes 
simultaneously, enlarging its symmetries.

Concerning the empirical observation of just three generations of fermions, 3-3-1 models 
offer a plausible explanation~
\cite{Singer:1980sw,Pisano:1991ee,Frampton:1992wt,Foot:1992rh,Montero:1992jk,Pleitez:1992xh,Foot:1994ym}. 
In these theoretical constructions the ${\rm SU}(2)_L\otimes {\rm U}(1)_Y$  symmetry group of the electroweak 
interactions is extended to ${\rm SU}(3)_L\otimes {\rm U}(1)_X$, in such a way that cancellation of all gauge 
anomalies involves necessarily all the three fermion generations. As it happens, there are different types of 3-3-1 models depending on the matter content fixed by
a parameter $\beta$ in the electric charge operator
\begin{equation}
Q = T_3 + \beta \ T_8 + X I\,,
\label{Qop} 
\end{equation}
where $T_3$ and $T_8$ are the diagonal generators of ${\rm SU}(3)_L$ built  as $T_a=\frac{\lambda_a}{2}$ 
from the Gell-Mann matrices $\lambda_a$, with $a =1,\,...,\,8$; and $X$ refers to the ${\rm U}(1)_X$ charge. 
Standard Model left-handed lepton fields take part in ${\rm SU}(3)_L$ triplets, $\psi_{iL}=(\nu_{i }\, e_{i}^-\, E_{i}^{q_E})^T_L$, having  $X_\psi=-\frac{1}{2}(1+\frac{\beta}{\sqrt3})$. The third components,  $E_{i}^{q_E}$, 
are new lepton fields with electric charge $q_E=-\frac{1}{2}(1+{\sqrt3}\beta)$. The particular choice 
$\beta=- \frac{1}{\sqrt 3}$ leads to models where the new leptons $E_{iL}$ do not carry an electric 
charge~\cite{Singer:1980sw,Montero:1992jk,Foot:1994ym,Hoang:1995vq}. Other well developed 
constructions have $\beta=-\sqrt3$ and, in this case, $E_{iL}$ carry one unit of electric charge so that it could 
be identified with the charged anti-leptons, i.e.,  $E_{iL}\equiv (l_{iR}^-)^c$~\cite{Pisano:1991ee,Frampton:1992wt}, or even represent new charged leptons~\cite{Pleitez:1992xh}.  For other possibilities, see~\cite{Diaz:2004fs}. 

For all models, the cancellation of gauge anomalies requires two anti-triplets (triplets) and one triplet (anti-triplet) 
of left-handed quarks when taking into account triplets (anti-triplets) of leptons. This feature leads to  flavor 
changing neutral currents (FCNC) mainly through a vector boson $Z'$ whose mass is related to the 
energy scale in which the ${\rm  SU}(3)_L\otimes {\rm U}(1)_X$ symmetry is broken down spontaneously to 
${\rm SU}(2)_L\otimes {\rm U}(1)_Y$. Bounds on the  $Z'$ boson mass have been obtained from the LHC 
data for the versions with $\beta=\pm1/\sqrt{3},\,-\sqrt{3}$ in Ref.~\cite{Salazar:2015gxa}, 
and recent analyses on FCNC have been performed in Refs.~\cite{Buras:2012dp,Queiroz:2016gif,Dong:2017ayu}. 

Many works have been published exploring the theoretical and phenomenological benefits 
of these models, showing that they are good candidates for describing new physics. 
For example, it is possible to include supersymmetry in such a context~\cite{Duong:1993zn,Montero:2000ng,Dong:2007qc,Huong:2012pg}, as well as to construct 
left-right extensions \cite{Dias:2010vt}, which have recently been subject of some 
studies~\cite{Reig:2016vtf,Reig:2016tuk,Franco:2016hip,Borah:2017inr}.

To implement spontaneous symmetry breaking, three or more scalar multiplets getting vacuum expectation value (vev) at the GeV-TeV scale  
have been considered in 3-3-1 models. As a consequence,  the scalar potential has many free parameters, 
being more complex than
e.g.\
the Two Higgs Doublet Model
(see Ref.\,\cite{Branco:2011iw} for a review). However, it has been shown that it is possible to break the symmetries down in 
some 3-3-1 models by taking into account two scalar triplets only~\cite{Ponce:2001jn,Ponce:2002sg,Dong:2006mg,Ferreira:2011hm,Dong:2017ayu}. These constructions are phenomenologically attractive once they have a simpler scalar potential, 
predicting only three Higgs bosons. With the introduction of effective operators, masses for all fermions 
can be generated~\cite{Ferreira:2011hm,Dong:2017ayu}. 

Our aim in this work is to develop a version of the 3-3-1 model with $\beta=\frac{1}{\sqrt3}$
distinct from its first proposals~\cite{Pleitez:1994pu,Ozer:1995xi} and
from other similar models focused on neutrino masses and 
mixing~\cite{Kitabayashi:2001jp,Ponce:2006au}.
We focus on this model because the versions with $\beta=\pm\sqrt{3}$~\cite{Pisano:1991ee,Frampton:1992wt,Pleitez:1992xh} 
become strongly interacting at an energy of few 
TeV~\cite{Frampton:1992wt,Ng:1992st,Frampton:2002st,Dias:2004dc,Dias:2004wk}. 
We comment on the case $\beta=-1/\sqrt{3}$ when relevant.
We show that a consistent symmetry breaking pattern is obtained for this model 
with only two scalar triplets getting vev.
Also, we 
present a simple mechanism where the low energy effective operators required to generate mass 
for some fermions arise after the integration of a supposedly heavy scalar triplet.
These effective operators share similarities with those in the Froggatt-Nielsen mechanism in the sense that they
generate more natural, less fine-tuned, masses for most of the fermions of the model, when compared to the mass generation in the Standard Model.
The scalar 
particle spectrum of the model is composed of just two neutral CP even scalars, with one of them 
directly identified with the discovered 125 GeV Higgs boson, plus a charged one and its antiparticle.
Also, five new vector 
bosons, $V^+$,  $V^-$,  $V^0$,  $V^{0\dagger}$,  and $Z'$, are predicted by the model. 
A study of the production signals at the LHC of the $Z'$ boson is performed. 

The paper is organized as follows: in Sec.~\ref{model}, the essential aspects of the model are presented, 
including symmetry breakdown, residual symmetries, and the particle spectra of scalar and vector bosons; 
Fermion masses are treated in Sec.~\ref{fmasses}; flavor changing interactions are analyzed in Sec. \ref{modFCNC}; in Sec.~\ref{uvcomp}, a simple UV-completion 
able to generate the needed effective operators for fermion masses is discussed; the 
$Z'$ boson phenomenology is presented in Sec.~\ref{Zpheno}; and our conclusions are given in 
Sec.~\ref{conclusions}.

\section{The model}
\label{model}

We focus on the model with $\beta= 1/\sqrt{3}$ in Eq.~\eqref{Qop}.  Therefore, the left-handed lepton fields form ${\rm SU}(3)_L$ triplets, with the right-handed lepton fields in ${\rm SU}(3)_L$ singlets, as follows 
\begin{eqnarray}
&& \psi_{i L} =  \left(\nu_i,e^{-}_i,E^{-}_i\right)_{L}^T
\ \sim \left( {\bf 1},{\bf 3}, -2/3 \right), \nonumber \\ 
&& \nu_{i  R}\ \sim \left({\bf 1},{\bf 1}, 0 \right), \,\,\,\,\,  
e^{\prime-}_{s R} \ \sim \left({\bf 1},{\bf 1}, -1 \right), 
\label{lepm}
\end{eqnarray} 
where $i = 1, 2, 3$, is the generation index, and $s=1,\,...,\, 6$, with $e^{\prime-}_{s R}\equiv (e^{-}_{iR},\,E^{-}_{i  R})$.
The numbers in parentheses refer to the field transformation properties under ${\rm SU}(3)_C$, ${\rm SU}(3)_L$, and ${\rm U}(1)_X$, respectively. We consider the right-handed neutrino fields, $\nu_{i  R}$, in order to generate small masses to the left-handed neutrinos through the usual seesaw mechanism. The fields $E^{-}_{i L}$,  required to complete the ${\rm SU}(3)_L$ representation, along with the right-handed components, give rise to three \textit{heavy leptons}. 

Given the above lepton multiplets, as first observed long ago, gauge anomalies are canceled
when the three families of quarks are included non-universally into
two anti-triplets and one triplet of ${\rm SU}(3)_L$ for the left-handed parts, 
and the corresponding right-handed fields assigned to singlets:
\begin{eqnarray}
& & Q_{a L} =  \left(d_a,-u_a,U_{a}\right)_{L}^T
\ \sim \left({\bf 3}, {\bf 3^*}, 1/3 \right),\nonumber \\
& & Q_{3 L} =  \left(u_3,d_3,D\right)_{L}^T
\ \sim \left({\bf 3},{\bf 3}, 0 \right),\nonumber \\
& & u_{m R}^\prime\ \sim \left({\bf 3},{\bf 1}, 2/3 \right), \,\,\,\,\,\,  d_{n R}^\prime \ \sim \left({\bf 3},{\bf 1}, -1/3 \right),
\label{quam}
\end{eqnarray}
where $a=1, 2$, $m=1,\dots,5$, $n = 1,\dots,4$, with  
$u_{mR}^\prime\equiv(u_{iR},\,U_{aR})$ and $d_{nR}^\prime\equiv(d_{iR},\,D_{R})$. Besides the quark fields of the SM, this model has two extra up-type quark fields, $U_{a}$,  and one down-type field, $D$. Such fields, as well as $E_i$, get their masses at the energy scale $w$, in 
which the ${\rm SU}(3)_L\otimes {\rm U}(1)_X$ is supposedly broken down to ${\rm SU}(2)_L\otimes {\rm U}(1)_Y$. Once that energy scale must be higher than the electroweak scale, i.e., $w>v=246$ GeV, it is natural for the new elementary fermions associated with those fields to be heavier than the standard ones. 

As we have already mentioned, the set of fields in Eqs.~\eqref{lepm} and~\eqref{quam} is such that the cancellation of gauge anomalies involves the three fermion generations. This contrasts with the SM where the cancellation of anomalies occurs in each family, independently. 

In principle, the choice of which generation of left-handed quark
is assigned to a triplet is arbitrary. But, the fact that not all left-handed quark multiplets have the 
same transformation properties leads to new sources of FCNC. This has been explored in various works 
considering different versions of 3-3-1 models. Constructions with the third generation transforming 
differently from the first two are less restricted by bounds of processes involving FCNC. 
We show in Sec.~\ref{modFCNC} that FCNC interactions are naturally 
suppressed in our model due to its peculiar mass generation mechanism for the fermions.

The following two scalar triplets realize the spontaneous breaking of the ${\rm SU}(3)_L\otimes {\rm U}(1)_X$ symmetry down to  $\mathrm{U}(1)_{Q} $ of the electromagnetic interactions:
\begin{eqnarray}
&  & \rho\equiv\left(\rho^{0}_1\,\,\rho^{-}_{2}\,\,\rho^{-}_{3}\right)^{T}
\sim\left(\mathbf{1},\mathbf{3,\,} -2/3\right),\nonumber \\
&  & \chi\equiv\left(\chi^{+}_{1}\,\,\chi^{0}_{2}\,\,\chi^{0}_{3}\right)^{T}
\sim\left(\mathbf{1},\mathbf{3,\,} 1/3\right).
\label{escm}
\end{eqnarray}
This is the minimal set of scalar fields that can perform the required symmetry breakdown. 

From the fermionic and the scalar multiplets in Eqs. (\ref{lepm}), (\ref{quam}), 
and (\ref{escm}), we write down the following Yukawa Lagrangian 
\begin{eqnarray}
-\label{Yuk}
\mathcal{L}_Y &=& h^E_{is} \overline{\psi_{i L}}~\chi\, e_{s R}^\prime + h^\nu_{ij} \overline{ \psi_{i L}}~ \rho \, \nu_{j R}+
\frac{1}{2}m_{ij} \overline{(\nu_{i R})^c} \nu_{j R}\nn\\
&+& h^U_{am}\overline{Q_{a L}}~ \chi^* u_{m R}^\prime+ h^d_{an}\overline{Q_{a L}}~ \rho^*d_{n R}^\prime+ 
f^u_{m}\overline{Q_{3 L}}~ \rho\, u_{m R}^\prime+f^D_{n}\overline{Q_{3 L}}~ \chi\,  d_{n R}^\prime +h.c.
\end{eqnarray}
where the complex coupling constants are such that: $ h^E_{is}$ is a $3\times 6$ matrix; $h^\nu_{ij}$, is a $3\times3$ matrix; $m_{ij}=m_i\delta_{ij}$ is a $3\times3$ diagonal matrix; $h^U_{am}$ and $h^d_{an}$ are  $2\times 5$ and $2\times 4$ matrices, respectively; $f^u_{m}$ and $f^D_{n}$ are $1\times 5$ and 
$1\times 4$ matrices, respectively. 

With only the two scalar triplets in Eq. (\ref{escm}), the most general renormalizable scalar potential is 
simply given by
\begin{eqnarray}
V(\chi,\rho)=\mu^{2}_{1}\rho^{\dagger}\rho+\mu
_{2}^{2}\chi^{\dagger}\chi
+\lambda _{1}(\rho^{\dagger}\rho)^{2}+\lambda _{2}(\chi^{\dagger}\chi)^{2} + \lambda _{3}(\chi^{\dagger}\chi)(\rho^{\dagger}\rho)+\lambda _{4}(\chi^{\dagger}\rho)(\rho^{\dagger}\chi)~.
\label{V}
\end{eqnarray}
We assume that the quadratic mass parameters, $\mu^2_{1,2}<0$, and the self-interaction coupling constants,  $\lambda_i$, $i=1,...4$, are such that the scalar fields will develop 
non-vanishing vevs, $\langle\chi\rangle$, $\langle\rho\rangle\neq 0$. 

\subsection{Symmetry breaking and residual symmetries}
\label{secResS}

Besides being invariant under the gauge symmetries 
${\rm SU}(3)_C \otimes {\rm SU}(3)_L \otimes {\rm U}(1)_X$, our model presents invariance under 
certain global symmetries that we will now describe.
When only renormalizable operators are taken into 
account, and right-handed neutrinos are not introduced, one can check that the 
Lagrangian is invariant under three extra global ${\rm U}(1)$ symmetries. Two of which can be 
taken as the Baryon and the Lepton number symmetries (${\rm U}(1)_B$ and ${\rm U}(1)_{Lep}$), 
while the other one, being associated with a $[{\rm SU}(3)_C]^2\times {\rm U}(1)$ anomaly, is a 
Peccei-Quinn-like symmetry (${\rm U}(1)_{PQ}$).\footnote{
We mean here that such a global symmetry is chiral and anomalous, 
but it will be broken explicitly, as discussed in Sec. \ref{fmasses}, implying that 
the Peccei-Quinn mechanism does not take place in our model. Implementations of the 
Peccei-Quinn symmetry for the strong CP problem in the 3-3-1 models can be found, for example, in Refs.~\cite{Pal:1994ba,Dias:2003iq}.}
The latter symmetry is the continuous version of the center of ${\rm SU}(3)_L$ 
in which every triplet, $\boldsymbol{3}$, carries unit charge while every anti-triplet, $\boldsymbol{3}^*$, carries the opposite charge. This is possible in our case due to the absence of any 
trilinear couplings of the form $\boldsymbol{3}\times\boldsymbol{3}\times\boldsymbol{3}$.
Therefore, when the scalar fields get vev, the ${\rm U}(1)_{PQ}$ symmetry will be broken, but its charges will be part of another remaining global symmetry as we show below. In Table~\ref{table1} the quantum numbers associated with the ${\rm U}(1)$ symmetries are presented for all the matter fields.
\begin{center}
	\begin{table}[h]
		\begin{tabular}{ |c|c|c|c|c|c|c|c|c|c| } 
			\hline
			$\#$ & $\chi$ & $\rho$ & $\psi_{i L}$ & ($e_{i R}$,\,$E_{i R}$) & 
			$\nu_{i R}$ & $Q_{a L}$ & $Q_{3 L}$ & $(d_{i R}, D_{3R})$ & $(u_{i R},U_{a R})$  
			\\\hline 
			$X$ & 1/3 & $-2/3$ & $-2/3$  & $-1$ & 0& 1/3  & 0 & $-1/3$ & 2/3 \\ \hline 
			$PQ$ & 1 & 1 & 1 & 0 & 0 & $-1$  & 1 & 0  & 0 \\  \hline 
			$B$ & 0 & 0 & 0 & 0 & 0 & 1/3  & 1/3 & 1/3  & 1/3 \\  \hline
			Lep & 0  & 0 & 1  & 1 & 1 & 0  & 0 & 0 & 0 \\  \hline 
		\end{tabular}
		\caption{Charges of the Abelian symmetries of the model.}
		\label{table1}
	\end{table}
\end{center}

Thus, our model is actually invariant under a larger symmetry group: 
${\rm SU}(3)_C \otimes {\rm SU}(3)_L \otimes {\rm U}(1)_X\otimes {\rm U}(1)_{PQ}\otimes {\rm U}(1)_B (\otimes {\rm U}(1)_{Lep})$ 
with the additional Abelian symmetries being global ones. We put ${\rm U}(1)_{Lep}$ 
in parentheses to call the reader's attention to the fact that such a symmetry 
is only present when the bare Majorana mass term for the $\nu_{iR}$ singlets in Eq. (\ref{Yuk}) is absent. 

Since scalar fields transform trivially under the ${\rm SU}(3)_C \otimes {\rm U}(1)_B(\otimes {\rm U}(1)_{Lep})$ 
subgroup, there is no way in which non-vanishing vevs of the scalar fields will trigger 
spontaneous symmetry breaking of such a structure. Consequently, we neglect this subgroup 
for now on and focus only on the groups affected by spontaneous symmetry breaking, {\it i.e.} 
${\rm SU}(3)_L\otimes {\rm U}(1)_X \otimes {\rm U}(1)_{PQ}$, containing ten independent generators.

We begin our analysis by noting that the electric charge operator in Eq.~\eqref{Qop} with $\beta=1/\sqrt{3}$ implies that the second and third component fields of any ${\rm SU}(3)_L$ triplet
or anti-triplet always have the same electric charge. This means that the charge operator is invariant by 
\be
\label{su2:rep}
	U^\dag QU=Q,
\ee
where $U$ can be an arbitrary ${\rm SU}(2)$ transformation in the 2-3 sub-block or any diagonal 
transformation.\footnote{Obviously, we can perform a more general reparametrization 
in ${\rm SU}(3)_L$ without physical consequence but the charge operator $Q$ would change.} It is the former group which is special to this case, and we denote this group as ${\rm SU}(2)_{\rm rep}$ where $\rm rep$ stands for reparametrization.	
The situation is different from many gauge extensions of the SM 
	in which there are no equal charge fields in the same multiplet, and the only transformations 
	that leave $Q$ invariant are the diagonal ones which include $Q$ itself, modulo Abelian factors. 
	This is also different from horizontal spaces that are present even before symmetry breaking, 
	such as in the Two-Higgs-doublet extension of the SM or the extension (UV completion) 
	of the current model with one more scalar triplet; see Sec.~\ref{uvcomp}.
	Reparametrization symmetry means that we can rotate all the fields of the theory by an 
	${\rm SU}(2)_{\rm rep}$ transformation, including the vev in Eq.\,\eqref{vevw}, 
	without affecting the physical content. The physical invariance is ensured because 
	${\rm SU}(2)_{\rm rep}$ is a subgroup of the original ${\rm SU}(3)_L$ global gauge group.

Therefore, without loss of generality we can consider that the minimum of the
potential in Eq.~(\ref{V}) is attained at the vevs
\begin{eqnarray} 
& & \langle\chi\rangle=\frac{1}{\sqrt{2}}\left(0, 0, w
\right)^T, \label{vevw}  \\ 
& & \langle\rho\rangle=\frac{1}{\sqrt{2}}\left(v,0,0
\right)^T.
\label{vevv}
\end{eqnarray}
If we had considered the more general vev $\langle\chi\rangle=\left(0, v^\prime, w^\prime\right)^T/\sqrt{2}$, 
the reparametrization symmetry in \eqref{su2:rep} would allow us to rotate $\langle\chi\rangle$ to the original form in Eq. (\ref{vevw}) 
without affecting $\langle\rho\rangle$ in Eq.~\eqref{vevv}. Consequently, the fields are also transformed so that 
Eqs.~\eqref{vevw} and \eqref{vevv} can be taken from the start. A direct consequence is that 
the vector bosons $V^\pm$ and $W^\pm$ do not mix at tree level. It has to be pointed out that the reparametrization 
symmetry applies to other models, like the 3-3-1 model defined by $\beta=-{1}/{\sqrt3}$, having a scalar triplet with two neutral components that could acquire vevs, so that we can make a rotation in order to have just one component with vev.  

Given that we are considering $w>v$, the spontaneous symmetry breaking induced by 
the vevs in Eqs. (\ref{vevw}) and (\ref{vevv}) happens in two stages: first with $\langle\chi\rangle$ 
realizing the breakdown 
\begin{equation}
{\rm SU}(3)_L \otimes {\rm U}(1)_X \otimes {\rm U}(1)_{PQ} \rightarrow
{\rm SU}(2)_L\otimes{\rm U}(1)_Y\otimes{\rm U}(1)_{PQ^{\prime}}\,,
\end{equation}
with the hypercharge given by $Y= T_8/{\sqrt3} + X$, and the charges of the global symmetry 
$PQ^{\prime}=3X-PQ$; and second with $\langle\rho\rangle$ realizing the breakdown 
\begin{equation}
\mathrm{SU}(2)_L\otimes\mathrm{U}(1)_Y \otimes\mathrm{U}(1)_{PQ^{\prime}}\rightarrow
\mathrm{U}(1)_Q\otimes \mathrm{U}(1)_{{\cal G}}
\end{equation}
with $Q$ the electric charge operator of Eq.~\eqref{Qop} and charges of a global 
symmetry given by\,
\footnote{If both neutral components of $\chi$ acquire a vev, {\it i.e.} 
$\sqrt{2}\langle \chi\rangle =(0,u,w)$, there still remains a conserved symmetry 
generated by
$ 
\mathcal{G}_\theta = 2(1 -2 \sin^2{\theta} ) T_3 +\sin{(2 \theta)} T_6 + (1-3 \sin^2{\theta}) X
- \ums{3}PQ~,
$	
	with $\tan{\theta}= u/w$
	for the case that $u,w$ are real and positive. If they were complex a more general 
		expression can be written reparametrized by ${\rm SU}(2)_{\rm rep}$.
}
\begin{equation}
\cG=2T_3+X-\ums{3}PQ\,.
\label{SG}
\end{equation}
It is easy to see that this operator is unbroken by the vevs of Eqs.\,(\ref{vevw}) and (\ref{vevv})
when we write it explicitly for $\chi$ and $\rho$:
$\mathcal{G}(\chi)=\diag(1,-1,0)$ and $\mathcal{G}(\rho)=\diag(0,-2,-1)$.
\footnote{
It is also clear that ${\rm U}(1)_{\mathcal{G}}$ would be generally broken if there is an additional scalar triplet that can acquire a vev in its second component.
	}
Thus, two independent generators out of the ten initial remain unbroken  after spontaneous symmetry breaking. The eight 
would-be Goldstone bosons associated with the broken generators are all absorbed to form 
the longitudinal degrees of freedom of the massive vector bosons: $Z$, $W^\pm$, 
$Z^\prime$, $V^{\pm}$, and the neutral non-Hermitian $V^0$ and $V^{0\dagger}$, with $\cG$ charges $0,\pm 2,0,\pm 1,-1$ and $+1$, respectively; see Eq.\,\eqref{G:gauge}. Consequently, from the twelve degrees of freedom contained in the two scalar triplets, four are left as physical scalar boson fields: two neutral CP even, $h$ and $H$, plus the charged ones $\varphi^\pm$. 
The mass spectra for the scalar bosons and vector bosons are shown in the next section. We anticipate some mass degeneracy from the conservation of $\mathcal{G}$.
For example, since $V^0$ and ${V^0}^\dag$ are the only neutral gauge bosons with $\cG$ charges
$\mp 1$, we expect that they remain mass degenerate and do not split into two neutral gauge bosons
with different masses.
This expectation is confirmed by the explicit calculation of the mass matrices; see
Sec.\,\ref{subsec:V}.

The ${\rm U}(1)_{\cal G}$ symmetry also has the property of being chiral for the second components 
of fermion triplets (anti-triplets) and their right-handed counterparts in singlets of ${\rm SU}(3)_L$. 
As a result the standard charged leptons, two up-type quarks and one down-type 
quark, are left massless even at the perturbative level, as 
pointed out in~\cite{Montero:2014uya}. In order to overcome this problem we introduce in 
Sec.~\ref{fmasses} dimension-5 operators which explicitly break the ${\rm U}(1)_{PQ}$ 
and, consequently, the ${\rm U}(1)_{\cal G}$ symmetry. As we will see, such operators can be generated 
at low energies in an UV-completed model with a heavy scalar triplet which is integrated out. 
In such a setting the $\cG$ charge is only broken by the soft breaking of PQ,
and thus it remains approximately conserved.

\subsection{Scalar bosons}
\label{sec:scalars}

In order to find the scalar field masses and corresponding physical states, let us first write the scalar triplets as
\be\label{scd}
\chi = \begin{pmatrix} \chi_1^{+} \\ \chi^0_2 \\
\frac{1}{\sqrt{2}}(w + S_3 + i A_3 )
\end{pmatrix}
~~\mbox{and}~~
\rho = \begin{pmatrix}
\frac{1}{\sqrt{2}}(v + S_1 + i A_1 ) \\
\rho_2^{-} \\ \rho_3^{-}
\end{pmatrix}
~,
\ee
where we have decomposed the neutral fields which acquire a non-vanishing vev into scalar and pseudo-scalar contributions, $S_i$ and $A_i$, respectively. In the approximation that the global charge $\cG$ is exactly conserved, we can expect from
its conservation that both $\chi_2^0$ and $\rho_2^-$ already have definite masses
(they are would-be Goldstone bosons) and
the possible pairs that can mix are $(\chi_1^+,\rho_3^+)$, $(S_1,S_3)$ and $(A_1,A_3)$,
assuming CP conservation.

The minimum condition for the potential leads to the constraint equations
\begin{eqnarray}
& & \mu^2_1+\lambda_1 v^2+\frac{1}{2}\lambda_3 w^2=0\nn \\
& & \mu^2_2+\lambda_2 w^2+\frac{1}{2}\lambda_3 v^2=0
\end{eqnarray}
from which the quadratic mass parameters $\mu_{1,2}^2$ can be eliminated.

The mass matrix, derived from the scalar potential, for the CP even scalars in the basis $(S_1,\,S_3)$ 
is
\begin{equation}
M^2_0=\left(
\begin{array}{cc}
2\lambda_1 v^2 & \lambda_3 v w    \\
\lambda_3 v w  & 2 \lambda_2 w^2
\end{array}
\right)
\label{mcpe}
\end{equation}
This leads to the quadratic mass eigenvalues 
\begin{eqnarray}
& & m^2_h = \lambda_1 v^2  +\lambda_2 w^2  
-\sqrt{ \lambda_3^2 v^2 w^2+\left(\lambda_2 w^2-\lambda_1 v^2\right)^2}~,\label{mh}\\
& & m^2_H = \lambda_1 v^2  +\lambda_2 w^2  
+\sqrt{ \lambda_3^2 v^2 w^2+\left(\lambda_2 w^2-\lambda_1 v^2\right)^2}~,\label{mH}
\end{eqnarray}
corresponding, respectively, to the mass eigenstates
\begin{eqnarray}
& & h = {\rm cos}\,\theta \,S_1 + {\rm sin}\,\theta \,S_3 \\ 
& & H = - {\rm sin}\,\theta \,S_1 + {\rm cos}\,\theta \,S_3 
\end{eqnarray}
where ${\rm tan}\,2\theta= {\lambda_3 v w}/({\lambda_2 w^2-\lambda_1 v^2})$. 
We identify $h$ as the state corresponding to the observed Higgs boson with
mass of $125$ GeV. In the limit $\theta\to 0$, the tree level couplings 
of $h$ to the electroweak vector bosons $W$ and $Z$ are the same as the 
Standard Model Higgs boson. 

The particle spectrum of the model does not contain CP odd neutral scalar fields. The pseudoscalar fields $A_1$ and $A_3$ are absorbed in the massive vector bosons $Z$ and $Z^\prime$. In particular, 
the complex field $\chi^0_2=(S_2+i A_2)/\sqrt{2}$ in the triplet $\chi$ 
does not get a mass term and plays the role of the Goldstone boson absorbed in the non-Hermitian neutral vector boson $V^0$ (both have $\cG$ charge $-1$).
This contrasts with the Two Higgs Doublet Models, which necessarily contain a neutral pseudoscalar in the particle spectrum. 

For the charged scalar fields, it can be seen that only $\rho_3^+$ and $\chi_1^+$ mix with each other 
so that their mass matrix, in the basis $(\rho_3^\pm,\,\chi_1^\pm)$, is 
	\begin{eqnarray}
	M^2_{\pm}=\frac{\lambda_4 }{2}\left(
	\begin{array}{cc}
	w^2 &  v w    \\
	v w  & v^2
	\end{array}
	\right),
	\label{mche}
	\end{eqnarray}
whose the nonzero eigenvalue 
	\be
	m^2_{\varphi^\pm} = \frac{\lambda_4 }{2}(v^2+w^2)~,\label{mch}
	\ee
	corresponds to the squared mass of a charged scalar state given by 
	\be
\label{varphi:pm}
	\varphi^\pm = \frac{1}{\sqrt{v^2+w^2}}( w\rho_3^\pm + v \chi_1^\pm ) .
	\ee
	The orthogonal eigenstates $G_{31}^\pm=(v\rho_3^\pm -w\chi_1^\pm)/\sqrt{v^2+w^2}$ 
	and $G_2^\pm=\rho^\pm_2$ are Goldstone bosons which are absorbed to form the 
	longitudinal components of the vector bosons $V^\pm$ and $W^\pm$. 

Thus, we see that four, from the initial twelve, degrees of freedom contained 
	in the two scalar triplets remain as the physical scalars $h$, 
	$H$, $\varphi^\pm$.  The other eight degrees of freedom become the Goldstone modes needed 
	to give mass to the vector bosons $W^\pm$, $Z$, $V^\pm$, $V^0$,  $V^{0\dagger}$, and $Z^\prime$. 
So $\rho$ contains predominantly the SM Higgs $h$ within the  
SU(2)$_L$ doublet and the heavy charged Higgs $\varphi^-$ in its third component while
$\chi$ contains predominantly the heavy Higgs $H$ in the third component and a small 
admixture of the $\varphi^+$ within the SU(2)$_L$.

\subsection{Vector bosons}
\label{subsec:V}

As usual, to determine the physical gauge bosons and their masses, we look at the 
covariant derivative terms for the scalar fields:
\begin{equation}
{\cal L} \supset 
(D_\mu\rho)^{\dagger}(D^\mu\rho) + (D_\mu\chi)^{\dagger}(D^\mu\chi),
\label{kinetic}
\end{equation}
in which the covariant derivative is defined as 
\begin{equation}
D_\mu  = \partial_\mu  - ig W^a_\mu T^a
-ig_X X B_\mu = \partial_\mu  - iP_\mu ,
\label{CovDer}
\end{equation}
where $T^a$, with $a=1,...,8$, are the ${\rm SU}(3)_L$ generators as defined in Eq.~\eqref{Qop}, and $X$ denotes the ${\rm U}(1)_X$ charge of the 
field on which $D_\mu$ acts; $g$, $g_{X}$ are the gauge coupling constants related 
to ${\rm SU}(3)_L$ and ${\rm U}(1)_X$, respectively. 
The gauge coupling constant $g$ is the same as in the Standard Model, since in 3-3-1 models 
the gauge group ${\rm SU}(2)_L$ is totally embedded in ${\rm SU}(3)_L$. 
	Additionally, the gauge coupling constants are related to the Standard Model 
	electroweak mixing angle $\theta _{W}$ according to 
	\begin{equation}
	t^2=\frac{g_X^2}{g^2}=\frac{{\rm sin}^2\theta _{W}}{1-\frac{4}{3}{\rm sin}^2\theta _{W}}.
	\end{equation}
Then, the $P_{\mu}$ matrix for $\boldsymbol{3}$ can be written as 
	\begin{equation}
	P_{\mu}=\frac{g}{2}\left(
	\begin{array}{ccc}
	W_{3\mu}+\frac{W_{8\mu}}{\sqrt{3}}+2t B_{\mu}  X & \sqrt{2}W^{+}_{\mu}  & \sqrt{2}V^{+}_{\mu}  \\
	\sqrt{2}W^{-}_{\mu}  & -W_{3\mu}+\frac{W_{8\mu}}{\sqrt{3}}+2t B_{\mu}   X & \sqrt{2}V^{0}_{\mu}  \\
	\sqrt{2}V^{-}_{\mu}  & \sqrt{2}V^{0\dagger}_{\mu}  & -\frac{2 W_{8\mu}}{\sqrt{3}}+2t B_{\mu}   X
	\end{array}
	\right),
	\label{xiii}
	\end{equation}
	where we have defined the following fields
	\begin{eqnarray}
	& & W^{+}_{\mu} = \frac{W_{1\mu}-iW_{2\mu}}{\sqrt{2}},\\
	& & V^{+}_{\mu} =\frac{ W_{4\mu} - iW_{5\mu}}{\sqrt{2}} ,\\ 
	& & V^{0}_\mu =\frac{W_{6\mu}- iW_{7\mu}}{\sqrt{2}}.
	\label{xiv}
	\end{eqnarray}
The $\cG$ charge carried by these gauge fields is given by $2T_3$ and yields
\be
\label{G:gauge}
\cG(P_\mu)=
	\begin{pmatrix}
	0 & ~2 &  \phantom{-}1 \cr
	  & ~0 & -1 \cr
	  &   &  \phantom{-}0 
	\end{pmatrix}\,.
\ee

The vector boson masses arise when the scalar fields in Eq.~(\ref{kinetic}) acquire vevs as in Eqs.~(\ref{vevw}) and (\ref{vevv}). The vector boson fields $W^{\pm}$, 
$V^{\pm}$, $V^0$, and $V^{0\dagger}$  get the following squared masses
\begin{eqnarray}
& & M^2_{W^\pm}=\frac{g^2 v^2}{4},\\  
& & M^2_{V^\pm}=\frac{g^2}{4}(v^2+w^2),\\
& & M^{2}_{V^0}=M^{2}_{{V^0}^\dag}= \frac{g^2}{4}w^{2}.
\end{eqnarray}
A direct consequence of breaking down the symmetries with just two scalar triplets is the tree level 
mass splitting prediction $M^2_{V^\pm}-M^2_{V^0}=M^2_W$. Another peculiarity of the model is that a novel sort of neutral current might occur involving $V^0$, $V^{0\dagger}$, since these vector bosons intermediate transitions between standard and new leptons with the same electric charge. 

The gauge boson fields $W_{3}$, $W_{8}$, and $B$ of the symmetry basis  
mix with each other leading to the mass matrix, in the basis $(W_{3},\,W_{8},\,B)$, 
\begin{equation}
M^{2}_0=\frac{g^{2}}{2}
\left(
\begin{array}{ccc}
\frac{v^{2}}{2}  & \frac{v^{2}}{2 \sqrt{3}} & -\frac{2 v^{2}}{3}  t \\ \frac{v^2}{2 \sqrt{3}} & \frac{\left(v^{2}+4 w^{2}\right)}{6}  &  \frac{-2\left( v^{2}+ w^{2}\right)}{3 \sqrt{3}}t \\ -\frac{2 v^{2}}{3}  t  & \frac{-2\left( v^{2}+ w^{2}\right)}{3 \sqrt{3}}t & \frac{2\left(4 v^{2}+w^{2}\right)}{9}  t^{2}
\end{array}
\right).
\label{xxi}
\end{equation}
The mass eigenstates from this matrix give the 
photon field, $A_\mu$, and two massive fields, $Z_{1\mu}$ and $Z_{2\mu}$,
\bea 
& & A_{\mu} = \frac{\sqrt{3}}{\sqrt{3+4t^2}}\left( t \  W^{3}_{\mu}+\frac{t}{\sqrt{3}} \ W^{8}_{\mu}+  B_{\mu}\right),\\
& & Z_{1\mu} = N_{Z_2}\left( -3 M^{2}_{Z_2}\ W^{3}_{\mu}+\sqrt{3} \left(3M^{2}_{Z_2}-g^{2} w^{2} \right)W^{8}_{\mu}+g^{2}w^{2} t \ B_{\mu}\right),\\
& & Z_{2\mu} =  N_{Z_1}\left(- 3 M^{2}_{Z_1}\ W^{3}_{\mu}+\sqrt{3} \left(3M^{2}_{Z_1}-g^{2} w^{2} \right)W^{8}_{\mu}+g^{2}w^{2} t \ B_{\mu}\right),
\eea 
where the normalization constants are
\begin{equation*}
N_{Z_2,Z_1}=\left[\left(g^{2}w^{2} t\right)^{2}+\left(3 M^{2}_{Z_2,Z_1}\right)^{2}+3\left(3M^{2}_{Z_2,Z_1}-g^{2}w^{2}\right)^{2}\right]^{-1/2}
\,.
\end{equation*}
The masses of the neutral vector bosons, $Z_{1\mu}$ and $Z_{2\mu}$, can be written as
\begin{equation}\label{Zmass}
M^2_{Z_2,Z_1} =\frac{g^2}{18} \left[(3 + 4 t^2) v^2 + (3 + t^2) w^2 \pm 
\sqrt{-9 (3 + 4 t^2) v^2 w^2 + ((3 + 4 t^2) v^2 + (3 + t^2) w^2)^2}\right]
\end{equation}
in such a way that the dominant contributions are
\begin{align}\label{Zpmass}
M^2_{Z_1}&= \frac{g^2v^2}{4\cos^2\theta_W}+{\cal O}\left(\frac{v^2}{w^2}\right),\nn\\
M^2_{Z_2} &= \frac{g^2\cos^2\theta_W w^2}{3-4\sin^2\theta_W}+{\cal O}\left(\frac{v^2}{w^2}\right).
\end{align}
And there is also the prediction, at tree level, that
\be
M^2_{Z_2}/M^2_{V^0} 
\approx 4\cos^2\theta_W/(3-4\sin^2\theta_W)\approx 1.48~,
\ee 
where we have used $\sin^2\theta_W \approx 0.231$. 

It is convenient for the discussion on  the FCNC in the model to express 
the mass eigenstates $Z_{2\mu}$ and $Z_{1\mu}$ as linear combinations of the fields 
$Z^\prime_\mu$ and $Z_\mu$, 
which result from the sequential symmetry breakdown SU(3)$_L\otimes$U(1)$_X$ and 
SU(2)$_L\otimes$U(1)$_Y$, respectively, 
\begin{equation}
\left(
\begin{array}{c}
Z_1   \\ 
Z_2 
\end{array}
\right)=
\left(
\begin{array}{cc}
{\rm cos}\,\theta  & -{\rm sin}\,\theta   \\ 
{\rm sin}\,\theta  & {\rm cos}\,\theta  
\end{array}
\right)
\left(
\begin{array}{c}
Z   \\ 
Z^\prime 
\end{array}
\right).
\label{zzl}
\end{equation}
The unitary matrix in Eq.~(\ref{zzl}) diagonalizes the $Z_\mu$--$Z^\prime_\mu$ mass matrix 
\begin{equation}
\left(
\begin{array}{cc}
M_Z^2  & M_{Z\,Z^\prime}^2   \\ 
M_{Z\,Z^\prime}^2  & M_{Z^\prime}^2  
\end{array} 
\right)\,,
\label{zzlmm}
\end{equation}
where
\begin{eqnarray}
& & M_Z^2=\frac{g^2 v^2}{4 \, \cos^2\theta_W}\,, \hspace{1 cm} 
M_{Z\,Z^\prime}^2 =-\frac{M_Z^2}{\sqrt{3-4 \, \sin^2\theta_W}}\,,
\nonumber\\
& & M_{Z^\prime}^2 = \frac{M_Z^2+g^2w^2\cos^2\theta_W}{{3-4 \, \sin^2\theta_W}}\,.
\end{eqnarray}
In terms of these elements, the angle $\theta$ is 
\begin{equation}
\tan (2\,\theta)=\frac{2\,M_{Z\,Z^\prime}^2}{M_{Z^\prime}^2-M_Z^2}.
\label{zzlma}
\end{equation}
For example, taking $M_{Z^\prime}\approx 4$~ TeV, the angle is then  
$\theta\approx {M_{Z\,Z^\prime}^2}/{M_{Z^\prime}^2}\approx - 4\times 10^{-4}$ so that  
$Z_{2\mu}\approx Z^\prime_\mu$.

Due to the fact that $M^2_{Z_1}$ also depends on the energy scale $w$ related to the 
breakdown of ${\rm SU}(3)_L\otimes {\rm U}(1)_X$ to ${\rm SU}(2)_L\otimes {\rm U}(1)_Y$, the model presents 
a deviation from the Standard Model $\rho_0$ pa\-ra\-me\-ter prediction  
$\rho_0=M_W^2/\cos^2\theta_W M_{Z^0}^2=1$, with $M_{Z^0}$ standing for the Standard Model 
$Z^0$ boson mass at tree level. Such a deviation, up to order 
$(v/w)^2\ll 1$ in $M^2_{Z_1}$, is given by 
\begin{equation}
\Delta\rho_0\equiv \frac{M_W^2}{\cos^2\theta_W M_{Z_1}^2}-1\approx 
\frac{(v/w)^2}{4\cos^4\theta_W},
\end{equation}
where $M_{Z_1}^2=M_{Z^0}^2+\delta M_Z^2$. The actual experimental data furnishes 
$\Delta \rho_0  \equiv \rho_0 -1 \lesssim 0.0006$~\cite{PDG}. 
Thus, if the tree level contribution is dominant over the radiative corrections 
we obtain the  lower-bound $w\geqslant 6.5$ TeV, taking into account the value 
$v=246$ GeV. For definiteness, we take $w=10\,\unit{TeV}$ which furnishes $M_{Z_2}\approx 4$ TeV, and $M_{V^\pm}\approx M_{V^0}\approx 3.2$ TeV, although in our phenomenological analysis in Sec.\,\ref{Zpheno} a lower value for the $Z'$ boson mass is also considered.

\section{Fermion masses}
\label{fmasses}

With the vevs $\langle\chi\rangle$ and $\langle\rho\rangle$ the Yukawa Lagrangian 
in Eq.~(\ref{Yuk}) leads to the mass matrix for neutrinos, in the basis 
$(\nu_{i L},{\nu_{i R}}^c)$, 
\be 
M^\nu = \frac{1}{2}\begin{pmatrix} 0 &  \frac{v}{\sqrt{2}}{\textbf{h}}^\nu \\ 
\frac{v}{\sqrt{2}}{\textbf{h}}^{\nu T}  & {\textbf{m}}\end{pmatrix}\,,
\label{numm}
\ee
where ${\textbf{h}}^\nu=(h^\nu_{ij})$.
We analogously denote by boldface the various Yukawa matrices appearing in Eq.\,\eqref{Yuk},
such as $\mathbf{h}^E,\mathbf{h}^U,\mathbf{h}^d,\mathbf{f}^u,\mathbf{f}^D$.
The mass matrix \eqref{numm} has the usual seesaw texture which 
generates masses at the sub-eV scale for the left-handed neutrinos assuming, 
for example, $h^\nu_{ij}$ of order one and $\mathbf{m}\sim 10^{14}$ GeV.

As we have already pointed out in Sec.~\ref{secResS}, 
due to the residual ${\rm U}(1)_{\cal G}$ symmetry, the Yukawa Lagrangian in 
Eq.~(\ref{Yuk}) is not sufficient for giving mass to all charged fermion fields. 
In order to overcome this problem, we consider the following dimension-5 
effective operators,\footnote{Effective operators like these have also been considered recently in a 
similar model in Ref.~\cite{Dong:2017ayu}.}
which can emerge from a simple ultraviolet completion shown in Sec.\,\ref{uvcomp},
\bea
-\mathcal{L} &\supset & \frac{y^e_{is}}{\Lambda}
\overline{\psi_{i L}} \,\chi^*\rho^* e_{s R}^\prime
+\frac{y^u_{a m}}{\Lambda} \overline{Q_{a L}}\, \chi\,\rho\, u_{m R}^\prime
+\frac{y^d_{ n}}{\Lambda} \overline{Q_{3 L}}\, \chi^*\rho^* d_{n R}^\prime +h.c.
\label{D5}
\eea 
The large mass scale is $\Lambda\gg w$ and
the matrices of coefficients $y^e_{is}$, $y^u_{a m}$ and $y^d_{n}$ have sizes
$3\times 6$, $2\times 5$ and $1\times 4$ respectively; they
are denoted by $\mathbf{y}^e,\mathbf{y}^u$ and $\mathbf{y}^d$ when the indices are suppressed.
A contraction of the ${\rm SU}(3)$ antisymmetric tensor $\varepsilon_{ijk}$ with the 
triplet fields in Eq.~(\ref{D5}) is implicit. 
These effective operators break explicitly the ${\rm U}(1)_{\cal G}$ symmetry allowing  
mass generation for the remaining charged fermion fields that are left massless 
when only Eq.~(\ref{Yuk}) is considered.

The $6\times 6$ charged lepton mass matrix, in the
basis $(e_{i },E_{i })_{L,R}$, has the form
\bea \label{clmm2}
\cM^l= 
\frac{1}{\sqrt{2}}
\begin{pmatrix} 
	\epsilon \, {\textbf{y}}^e \\  w\, {\textbf{h}^E} 
\end{pmatrix}
=
\begin{pmatrix} 
	M_e & M_{eE} \\ 0_{3\times 3} &  M_E 
\end{pmatrix}
\,,
\eea 
where $M_e,M_{eE},M_E$ are all $3\times 3$ matrices and the first two matrices are 
hierarchically suppressed by $\epsilon= {v w}/{\sqrt{2}\Lambda}\ll v$. 
We can choose the lower left block to vanish without loss of generality ---and without
disrupting the natural hierarchy between the first three rows ($\sim\eps$) and the last 
three ($\sim w$)--- 
by rotating appropriately in $e'_{sR}$ whose rotation matrix is unphysical as all fields are 
singlets of the gauge group.
In this form, $M_e$ already approximately represents the mass matrix for the charged 
leptons $l_\alpha$, $\alpha=e,\mu,\tau$, of the SM and $M_E$ represents the mass matrix 
for the heavy charged leptons $\cE_i$, $i=1,2,3$.
By further exploring the freedom to rotate $\psi_{iL}$ we could make either $M_e$ or $M_E$ diagonal.
There is a mixing among the $l_{\alpha L}$ and ${\cal{E}}_{iL}$ controlled by 
the entry $m_{eE}$ and has magnitude suppressed by $M_{eE}/M_E\sim \eps/w$.
In the limit $\epsilon\rightarrow 0$ ($\Lambda\rightarrow\infty$) the eigenstates $l_i$ 
become massless as a result of the ${\rm U}(1)_{\cal G}$ symmetry restoration. Therefore, 
it is natural that the leptons $l_\alpha$ are lighter than ${\cal E}_i$. 

The above scenario associates the mass of the known charged leptons to the 
energy scale $\epsilon$ which is derived from the electroweak scale $v$ times 
a suppression factor $w/\Lambda$.
For example, for $w=10$ TeV and $\eps\sim m_\tau \sim 1$ GeV, we have $\Lambda\sim 10^3$ TeV. Although there is still 
an unexplained fine tuning of order $10^{-3}$ for the electron mass relative to 
the scale $\epsilon$, this situation contrasts with the Standard Model where a 
tuning of order $10^{-5}$ relative to the electroweak scale is required. 

For the up-type quark mass matrix, in the basis $(u_1,u_2,u_3,U_1,U_2)^T_{L,R}$, 
we obtain similarly
\bea
\label{MM:u}
\cM^{u} =  \frac{1}{\sqrt{2}}
\begin{pmatrix}  
	-\epsilon \textbf{y}^u  \\ 
	v \textbf{f}^u \\ 
	w \textbf{h}^U
\end{pmatrix}
=
\mtrx{M_u & M_{uU} \cr 0_{2\times 3} & M_U}
\,,
\label{uqmm}
\eea 
after an appropriate redefinition of $u'_{mR}$, $m=1,\dots,5$; $M_u,M_{uU},M_U$ are matrices of sizes 
$3\times3$, $3\times 2$, and $2\times 2$, respectively.
By also rotating $Q_{aL}$ we can choose 
$M_U=\diag(m_{{\cal U}_1},m_{{\cal U}_2})$ as diagonal,
whose values of order 
$w$ correspond to the heavy quarks ${\cal U}_1,{\cal U}_2$ of charge $2/3$.
Analogously, $M_u$ corresponds to the mass matrix of the up-type quarks of the SM, 
$(u,c,t)$.
The large separation in energy among the sets of rows in \eqref{MM:u} naturally 
suppresses the mixing between states with hierarchically different 
masses\,\cite{delAguila:1982fs}.
The mixing between the heavy quarks $\cU_{aL}$ and the SM quarks (left-handed)
are controlled by the entry $M_{uU}$ and is at most $M_{uU}/M_U \sim 
v/w\sim 10^{-2}$ for $t_L$ and at most of order $\eps/w\sim 10^{-4}$ for $(u_L,c_L)$.
The entries of $M_u$ themselves have a natural hierarchy of $\eps/v\sim 10^{-2}$ between the first two
rows and the third.
By conveniently rotating the right-handed components, we can write the mass matrix for the
SM up-type quarks,
\eq{
\label{Mu:SM}
M_u=
\frac{1}{\sqrt{2}}
\begin{pmatrix}  
	\epsilon y_i^u  \\ 
	v f_i^u \\ 
\end{pmatrix}
=\mtrx{\tM_u & \tM_{ut}\cr 0_{1\times 2} & m_t}\,,
}
where $m_t$ is the top mass of order $v$ and $\tM_u,\tM_{ut}$ are of order $\eps$ or
smaller. The mass matrix $\tM_u$ is naturally of order $\eps\sim 1\,\unit{GeV}$ and 
gives masses for $(u,c)$.
We could have chosen $\tM_u$ to be diagonal instead of $M_U$ from the rotation on 
$Q_{aL}$.
The mixing between $t_L$ and $(u_L,c_L)$ is naturally suppressed by
$\tM_{ut}/m_t\sim \eps/v\sim 10^{-2}$.

For the down-type quarks, in the basis $(d_1,d_2,d_3,D)^T_{L,R}$, we have the mass matrix
\bea 
\cM^d=  \frac{1}{\sqrt{2}}
\begin{pmatrix}  
	v  \textbf{h}^d  \\ 
	\epsilon \textbf{y}^d \\ 
	w \textbf{f}^D
\end{pmatrix}
=
\mtrx{M_d & M_{dD} \cr 0_{1\times 3} & M_D}
\,,
\label{dqmm}
\eea
after appropriate rotation in $d'_{nR}$, $n=1,\dots,4$.
The mass $M_D$ of order $w$ corresponds to a new heavy quark $\cD$ while SM quarks $(d,s,b)$ 
have a mass matrix given approximately by $M_d$.
The mixing between $\cD_L$ and the $(d_L,s_L,b_L)$ are naturally suppressed by at least
$M_{dD}/M_D\sim v/w\sim 10^{-2}$.
In fact, for the down-type quarks, we do not obtain a natural hierarchy between the 
first
two families and the third family.
We obtain a natural hierarchy if $h^d_{an}$ is not of order one but suppressed by
\eq{
\label{h:d}
vh^d_{an}=\eps'\bh^d_{an} \sim 6\times 10^{-4}\,v\bh^d_{an}\,,
}
with $\bh^d_{an}$ of order one and $\eps'\sim m_s\sim 0.1\,\unit{GeV}$.
This suppression further decreases the mixing between $\cD_L$ and $b_L$ to $\eps/w\sim 10^{-4}$
and one order of magnitude smaller for the mixing with $(d_L,s_L)$.
We show in the following a possible mechanism responsible for this further suppression.
Assuming this hierarchy for the moment, we obtain the mass matrix for $(d,s,b)$:
\eq{
\label{Md:SM}
M_d=
\frac{1}{\sqrt{2}}
\begin{pmatrix}  
	\eps' \bh_{ai}^d  \\ 
	\epsilon y_{i}^d 
\end{pmatrix}
=\mtrx{\tM_{d} & \tM_{db}\cr 0_{2\times 1}& m_b}
\,,
}
where we have used appropriate rotations on $d'_{iR}$, $i=1,2,3$.
We can see that $m_b$ is naturally of order $\eps\sim 1\,\unit{GeV}$ and $\tM_d$ ---which
yields the masses for $(d,s)$--- is naturally of order $\eps'\sim m_s$.
The mixing between $b_L$ and $(d_L,s_L)$ is naturally suppressed by
$\tM_{db}/m_b\sim m_s/m_b\sim 0.02$.

The necessary suppression in $\textbf{h}^d$  could arise if the operator $h^d_{an}\overline{Q_{a L}}\,\rho^*d_{n R}^\prime$ in Eq.~(\ref{Yuk}) is in fact absent at tree-level but results from 
an effective higher order operator involving a new singlet scalar $\varphi$ at a very high energy 
as
\begin{equation}
\bh^d_{an}\frac{\varphi}{\Lambda^\prime}\overline{Q_{a L}}\,\rho^*d_{n R}^\prime + h.c.
\rightarrow 
\bh^d_{an}\frac{\langle\varphi\rangle}{\Lambda^\prime}
\overline{Q_{a L}}\,\rho^*d_{n R}^\prime + h.c..
\end{equation}
Thus, we would have effectively that 
$h^d_{an}=\bh^d_{an}{\langle\varphi\rangle}/{\Lambda^\prime}$, where 
${\langle\varphi\rangle}/{\Lambda'}\sim \eps'/v\sim 6\times 10^{-4}\ll 1$.
The absence of the tree-level term $\overline{Q_{a L}}\rho^*d'_{n R}$ can be arranged by
introducing a $\ZZ_2$ symmetry under which only $\varphi,Q_{aL},u_{aR},U_{aR}$ are odd.
One of the up-type quarks, $u_{3R}$, is kept even so that the top mass is still generated
correctly by Eq.~\eqref{Yuk}.
The direct interaction terms involving the 
bilinear forms  $\overline{{Q}_{aL}}u_{3R}$, 
$\overline{{Q}_{3L}} u_{aR}$ and $\overline{{Q}_{3L}} U_{aR}$ will be absent but effectively induced by the replacements $\overline{{Q}_{aL}}\to \varphi/\Lambda'\overline{{Q}_{aL}}$,  
$u_{aR}\to u_{aR}\,\varphi/\Lambda'$, and $U_{aR}\to U_{aR}\,\varphi/\Lambda'$, so that 
the mixing between $t_L$ and the heavy $\cU_{aL}$ or the lighter $(u_L,c_L)$
will be further suppressed by $\eps'/v\sim 10^{-3}$ compared to the estimates discussed above.
This property renders the top quark essentially unmixed with the rest.

Therefore, our minimal mechanism of breaking the 3-3-1 symmetry by the use of just two triplets,
together with the $\ZZ_2$ symmetry above,
correctly displays the qualitative hierarchy between the masses for the third
family quarks and those of the first two families.
Comparing the mass matrices in Eqs.~\eqref{Mu:SM} and \eqref{Md:SM}, it is clear that we have enough
freedom to obtain the correct masses for the SM quarks and the correct Cabibbo-Kobayashi-Maskawa (CKM) mixing matrix in a quantitative way.
We can write
\eq{
M_d=(V^d_L)^\dag \diag(m_d,m_s,m_b)\,,
\quad
M_u=(V^u_L)^\dag \diag(m_u,m_c,m_t)\,,
}
already discarding the unobservable rotation matrices for the right-handed quarks.
The CKM matrix is fixed by 
\eq{
V_{\rm CKM}=(V^u_L)^\dag V^d_L\,,
}
so that $V^d_L$ can be considered free, while $V^u_L$ is fixed by 
$V_{\rm CKM}$ and $V^d_L$.
From the discussion above, $V^u_L$ is essentially block-diagonal with the third family
practically decoupled from the rest. 
Another possibility to naturally suppress the coupling $h^d_{an}$ would be to implement 
the Froggatt-Nielsen mechanism~\cite{Froggatt:1978nt} (see also~\cite{Huitu:2017ukq} 
for a proposal along this line in a 3-3-1 model).
We leave this question for future investigations.

We briefly comment on the model with $\beta=-1/\sqrt{3}$ where the heavy quark content
is inverted ---we would have two heavy quarks $D_{a}$ of charge $-1/3$ and just one quark $U$ of charge
$2/3$--- and the heavy charged leptons $E_i$ are replaced by neutral leptons $N_i$ that could
participate in the mass generation mechanism for light neutrinos.
The natural hierarchy in the up-type and down-type quark sectors would be very 
similar and
the implementation of the $\ZZ_2$ symmetry is analogous.

\section{Suppressed flavor changing currents}
\label{modFCNC}

We consider first the FCNC interactions mediated by scalars.
To that end it is instructive to consider first the breaking of $SU(3)_L\otimes U(1)_X$ to the SM gauge group by $\aver{\chi}$ and rewrite
\eq{
\chi=\mtrx{-\phi_1 \cr \chi_3^0}
\,,
\quad
\rho=\mtrx{\tphi_2 \cr \rho_3^-}
\,,
}
where $\phi_1,\phi_2$ are $SU(2)_L$ doublets of $Y=1/2$.
At this stage, only $\sqrt{2}\re\chi_3^0$ acquires a vev $w\sim 10\,\unit{TeV}$
and $\sqrt{2}\re(\chi_3^0)$ and $\rho_3^-$ will be heavy SM singlet scalars of $Y=0$ and $Y=-1$ respectively.
The scalar $\sqrt{2}\im\chi_3^0$ and the doublet $\phi_1$ will be absent in the unitary gauge
because they will be absorbed by the gauge bosons $Z'_\mu$ and $(V^{+}_\mu,V^0_\mu)^T$.\,%
\footnote{The charged component of $\phi_1$ will have a small admixture with $\rho_3^+$
after Electroweak Symmetry Breaking (EWSB); see Eq.\,\eqref{varphi:pm}.
}
See Sec.\,\ref{sec:scalars} for their composition after EWSB.
When taking into account the effective operators in \eqref{D5}, it is convenient to write
\eq{
\frac{1}{\Lambda}\chi^*\times\rho^*=
\mtrx{\phi_3\cr \frac{\phi_1^\dag\phi_2}{\Lambda}}
\,,
}
where
\eq{
\phi_3=\frac{1}{\Lambda}(\chi_3^{0*}\phi_2-\rho_3^+\tphi_1)
\sim \frac{\re{\chi_3^0}}{\Lambda}\phi_2\,,
}
is a SM effective Higgs doublet. The dominant contribution coming from \eqref{D5} at this stage
will be
\eq{
\phi_3\to \frac{\aver{\chi_3^0}}{\Lambda}\phi_2
=\frac{\eps}{v}\phi_2
\sim 10^{-2}\phi_2\,.
}

We also separate the quark triplets in Eq.\,\eqref{quam} and lepton triplets in Eq.\,\eqref{lepm}
into SM doublets and singlets as
\eq{
Q_{aL}
=\mtrx{i\sigma_2 q_{aL}\cr U_{aL}}
\,,
\quad
Q_{3L}=\mtrx{q_{3L}\cr D_{L}}
\,,
\quad
\psi_{iL}=\mtrx{l_{iL}\cr E_{iL}}
\,,
}
where $q_{aL}$ is the usual quark doublet of family $a=1,2$, $\sigma_2$ is one of
the Pauli matrices, and $l_{iL}$ are the usual lepton doublets.

At this stage of breaking, the model is equivalent to the SM with additional
heavy singlet vector-like quarks (VLQ) $D,U_{1,2}$, singlet vector-like leptons $E_i$ and heavy singlet scalars $\sqrt{2}\re(\chi_3^0),\rho_3^{\pm}$, with the additional heavy gauge bosons.
The Yukawa interactions in \eqref{Yuk} and \eqref{D5} that only involve
the light doublet $\phi_2$ are
\eqali{
\label{L:yukawa:2+1}
-\lag_0&=
\bar{q}_{aL}\phi_2\ums[\sqrt{2}]{\eps'}\big[(M_d)_{ai}d_{iR}+(M_{dD})_{a}D_R\big]
+
\bar{q}_{3L}\phi_3\ums[\sqrt{2}]{\eps} \big[m_b d_{3R} + (M_{dD})_{3}D_R\big]
\cr
&~~
+\ \bar{q}_{aL}\tphi_3\ums[\sqrt{2}]{\eps} \big[(M_u)_{ai}u_{iR}+(M_{uU})_{ab}U_{bR})\big]
+\bar{q}_{3L}\tphi_2\ums[\sqrt{2}]{v}\big[m_t u_{3R} + (M_{uU})_{3b}U_{bR}\big] 
\cr
&~~
+h.c.
}
See appendix \ref{ap:op} for other interactions involving heavy particles.
Note that $d_{3R}\approx b_R$, $u_{3R}\approx t_R$ and $q_{3L}\approx (t_L,b_L)^\tp$
are almost the mass eigenstates except for suppressed mixing with the lighter quarks or 
the heavier $\cD,\cU_a$.
With only one effective Higgs doublet, there is natural flavor conservation and no flavor changing
neutral interactions mediated by scalars\,\cite{Glashow:1976nt}, except for the ones induced
by the small mixing between SM and heavy quarks\,\cite{delAguila:1982fs}.
This contrasts with the usual mechanism for breaking the 3-3-1 symmetry involving
three $\mathrm{SU(3)}_L$ triplets: usually there are two light Higgs doublets which induce
suppressed but nonvanishing neutral flavor changing interactions at tree level\,\cite{Liu:1994rx}.

We can consider now the currents coupling with the gauge bosons.
There are two types of FCNC interactions for SM fermions:
(i) the ones coming from the mixing between the third family of quarks and the first two families
and (ii) the ones coming from the mixing between heavy fermions and the SM fermions.
The first type (i)
inevitably appears in all 3-3-1 models because one of the quark families is treated differently
by the $\mathrm{SU(3)}_L$. Treating the third family differently is the only option if we want to
avoid unrealistic flavor changing contributions in the mass differences of the
$K^0,D^0$ and $B^0$ systems for a $Z_2^\mu$ of mass no larger than a few 
TeV\,\cite{Liu:1994rx}.
Sufficiently suppressed mixing of the third family with the lighter ones thus leads 
to experimentally allowed flavor changing contributions.
In our model, however, because of the different mass generation mechanism, such a mixing
is already \textit{naturally} suppressed at least at the level of $\eps/v\sim 10^{-2}$ for 
the up-type quarks (further suppressed with the $\ZZ_2$ implementation) and at least
$m_s/m_b\sim 10^{-2}$ for the down-type quarks.
The flavor changing interactions of type (ii) are well known to be naturally suppressed by the
hierarchy of SM fermions and heavy fermions\,\cite{delAguila:1982fs}.
We discuss these suppressed interactions in more detail in the following.

The fermionic currents couple with a vector boson $V^\mu$ as
\eq{
\lag\supset -V_\mu J^\mu_V\,.
}
For the neutral gauge boson $V=Z'$, we have
\eqali{
J^\mu_{Z'}&=
g_{Z'}
\bar{\Psi}\gamma^\mu\left[s_W^2 Y-\sqrt{3}c_W^2 T_8\right]\Psi
}
where $g_{Z'}\equiv \frac{g}{\sqrt{3}c_W \sqrt{1-4s^2_W/3}}$ and 
$\Psi$ is the collection of all the fermions in the symmetry basis.
The flavor changing piece can be extracted as
\eqali{
\label{FCNC:Z'}
g_{Z'}^{-1}J^\mu_{Z'}\big|_{\rm FCNC}&=
(-c^2_W)\bar{u}_{3L}\gamma^\mu u_{3L}
+(-c^2_W)\bar{d}_{3L}\gamma^\mu d_{3L}
\cr&~~
+\ (-\ums[3]{2}+2s_W^2)\bar{U}_{aL}\gamma^\mu U_{aL}
+(\ums{2}-s_W^2)\bar{D}_{L}\gamma^\mu D_L
\cr&~~
+\ (c_{2W}+\ums{2})\bar{E}_{iL}\gamma^\mu E_{iL}
\,,
}
where $c_{2W}\equiv\cos 2\theta_W$, and we have subtracted the flavor universal part common to the first two families of quarks and three family of leptons.
Small mixing terms appear when we go to the basis of physical fields of definite masses.
We can clearly see in \eqref{FCNC:Z'} the two types of FCNC discussed above:
the first line induces interactions of type (i) while the rest induces the type (ii) 
currents.
For type (i), only the mixing between the third family and the first two families will be observable.
This mixing is naturally suppressed by $\eps/v\sim m_s/m_b\sim 10^{-2}$ in our model
which is typically below the current limits coming from meson mixing\,\cite{Promberger:2007py}.

As the interaction with the $Z$ boson is identical to the ones in the SM for the usual
quarks and leptons, FCNCs are only of type (ii):
\eqali{
\label{FCNC:Z}
g_{Z}^{-1}J^\mu_{Z}\big|_{\rm FCNC}&=
-\ums{2}\bar{U}_{aL}\gamma^\mu U_{aL}
+\ums{2}\bar{D}_{L}\gamma^\mu D_{L}
+\ums{2}\bar{E}_{iL}\gamma^\mu E_{iL}
\,,
}
where $g_Z\equiv g/c_W$.
We have again subtracted the family universal contributions from the first families
of quarks and leptons.
The coupling with $W$ is the same as in the SM in the symmetry basis and small non-unitary effects
appear through mixing between heavy and SM fermions.
Flavor changing interactions are constrained by the search for singlet VLQs at the LHC\,\cite{Araque:2016jrb}
and by indirect constraints coming from precision electroweak observables and from the Large Electron-Positron Collider (LEP) \,\cite{Aguilar-Saavedra:2013qpa}.
The former constrains the masses to be above around 1 TeV and the latter constrains
the mixing angle between the heavy quarks and the third family to be less than
0.04 for the down-type quarks and 0.14 for the up-type quarks for heavy quarks of 1 TeV.
So our heavy quarks of masses at the scale $w\sim 10\,\unit{TeV}$ and mixing angle of
less than $10^{-2}$ are not currently observable.
If the heavy quark masses are lowered to few TeV, and the $\ZZ_2$ that decouples the top
is present, the dominant channel for $\cU_a$ will be $\cU_a\to Wb+X$, as
$\cU_a\to ht+X$ and $\cU_a\to Zt+X$ will be negligible.
For $\cD$, the channels $\cD\to hb+X$ and $\cD\to Wt+X$ are similarly important.
The constraints for singlet vector-like leptons are much more relaxed.

For completeness, we also collect the interactions with the heavy gauge bosons $V^0,V^+$:
\eq{
-\lag_V=
\frac{g}{\sqrt{2}}\left[
-\bar{q}_{aL}\gamma^\mu\mtrx{V^{0\dag}_\mu\cr -V^-_\mu}U_{aL}
+\bar{q}_{3L}\gamma^\mu\mtrx{V^+_\mu\cr V^0_\mu}D_L
+\bar{l}_{iL}\gamma^\mu\mtrx{V^+_\mu\cr V^0_\mu}E_{iL}
\right]
+h.c.
\,,
}
where the heavy gauge bosons $(V^+_\mu,V^0_\mu)^\tp$ have the same gauge quantum numbers
as the SM Higgs doublet.
These gauge bosons lie at the scale $w$ and interactions with two SM fermions are suppressed
by the heavy-light mixing.

\section{An ultraviolet completion of the model}
\label{uvcomp}

We show in this section a simple ultraviolet completion of the model 
allowing for the generation of the effective operators in Eq.~(\ref{D5}). 
In order to achieve that, we add a scalar field  
$\eta\sim\left(\mathbf{1},\mathbf{3,\,} 1/3\right)$ which transforms in the same way as $\chi$ under ${\rm SU}(3)_C\otimes {\rm SU}(3)_L\otimes {\rm U}(1)_X$. 
We assume that $\eta$ has a mass $M\gg w,v$ much larger than the rest of the 
other fields in the model. Thus, as we describe below, $\eta$ can be integrated out so that the 
remaining effective theory is exactly the model studied above with 
the dimension-5 effective operators given in (\ref{D5}). 

With the introduction of $\eta$, the total scalar potential is 
\be 
V_T(\eta,\chi,\rho)= V(\chi,\rho)+V_\eta,
\ee 
where $V(\chi,\rho)$ is given in Eq.~(\ref{V}) and 
\begin{eqnarray}
V_\eta &=& M^{2}\eta^{\dagger}\eta+\left(\lambda _{5}(\eta^{\dagger}\eta) 
+\lambda_{6}(\rho^{\dagger}\rho) +\lambda_{7}(\chi^{\dagger}\chi)\right)
(\eta^{\dagger}\eta) +\lambda _{8}(\eta^{\dagger}\rho)(\rho^{\dagger}\eta)
+\lambda _{9}(\eta^{\dagger}\chi)(\chi^{\dagger}\eta)\nn\\
&-&\left[f\,\eta\rho\chi - \lambda_{10}(\eta^\dagger \rho)(\rho^\dagger \chi) 
-\left(\lambda_{11} \eta^\dagger \chi+\lambda_{12}\eta^\dagger \eta
+\lambda_{13}\chi^\dagger \chi+\lambda_{14}\rho^\dagger \rho\right)(\eta^\dagger \chi) 
+ h.c.\right],
\label{newV}
\end{eqnarray}
where $M^2>0$ is the quadratic mass for $\eta$; with the coupling constant $f$, which we 
take as being real, having dimension of mass; and the $\lambda$'s ($<4\pi$) are the 
usual scalar field perturbative 
self-interaction coupling constants.
Also, we consider a basis $(\chi,\,\eta)$ in which the bilinear terms in these fields are diagonal. 
In fact, bilinear terms, such as $\mu_3^2\, \eta^\dagger\chi$, can be eliminated through a rotation 
to the diagonal basis, implying effectively a change on the original quadratic 
mass parameters along with a redefinition of the quartic coupling constants.

We can see that the approximate conservation of the global charge $\cG$ in this UV completion is guaranteed
by the fact that the breaking is induced solely by a soft breaking of the PQ symmetry
through the $f$ term in \eqref{newV}.
Therefore the breaking effects are all proportional to the breaking parameter $f$ even
if we consider radiative corrections, and this fact justifies the approximate conservation
of $\cG$ at low energies.

The Yukawa interactions involving $\eta$ are similar to those for $\chi$ 
in Eq.~(\ref{Yuk}), adding to such an equation the terms 
\bea   
\mathcal{L}_Y &\supset & y^e_{is} \overline{\psi_{i L}}~\eta\, e_{s R}^\prime 
+  y^u_{am}\overline{Q_{a L}}~ \eta^* u_{m R}^\prime 
+ y^d_{n}\overline{Q_{3 L}}~ \eta\,  d_{n R}^\prime +h.c..
\label{yeta}
\eea 

Assuming $M\gg|f|\gtrsim w$, at low energies $\eta$ is effectively given by
\be 
\eta \approx \frac{f}{M^2} \rho^* \chi^* +\cdots,
\label{etaeff}
\ee 
where the ellipsis stands for operators which are even more suppressed by $M$.%
\footnote{There are other terms of the same order in $1/M^2$ that correct
the contribution of $\chi$ but only \eqref{etaeff} leads to a vev in a direction orthogonal to
$\aver{\chi}$ and $\aver{\rho}$. }
Replacing this last expression for $\eta$ 
in Eq.~(\ref{yeta}), we get the effective operators in Eq.~(\ref{D5}) with 
the identification $\Lambda = M^2/f$. Moreover, there exist corrections to tree level parameters that shift the couplings $\lambda_3\rightarrow {\lambda}_3-|f|^2/M^2$ and 
$\lambda_4\rightarrow {\lambda}_4+|f|^2/M^2$. As an example, the value $\Lambda=983$ TeV 
can be achieved with $M\approx 10^5$ GeV and $f\approx 10^4$ GeV. We see that 
for $|\lambda_{3,4}|$ of order unity, the correction $|f|^2/M^2\sim 10^{-2}$ 
does not have a significant impact on those couplings.

\section{Phenomenology of the $Z^\prime$ boson}
\label{Zpheno}

In this section, we present some results involving the new neutral vector boson  within the 
context of the LHC at the 14 TeV energy regime. By the reason that the mixing angle $\theta$ 
in Eq.~(\ref{zzlma}) between $Z$ and $Z^{\prime}$ is small for $w\gg v$,  
we have $Z_2\approx Z^{\prime}$. 
Thus, we consider in the following analysis the new vector boson as being $Z^{\prime}$ 
and its couplings with fermions. 
Constraints on the $Z^{\prime}$ mass coming from FCNCs can be strongly dependent
on the specific model of choice. In particular, for 3-3-1 models with $\beta$ =$\pm 1/\sqrt{3}$, it has been shown that by choosing
either the first or the third quark family to transform differently from the others leads to different constraints on the $w$ scale and, consequently, on the new gauge boson masses \cite{Dong:2017ayu}. For other versions, a
lower-bound of 3 TeV for the $Z^{\prime}$ mass has been found \cite{BUR}.

In addition, previous studies concerning the $Z^{\prime}$ branching ratios for the $\beta= \pm \sqrt{3}$ versions have identified a leptophobic character of such a neutral gauge boson \cite{ZPHOBIC}. In our case, within a scenario where the exotic masses are just above 1 TeV, the $Z^{\prime}$ branching ratios are divided into Br[$Z^\prime \rightarrow \nu \bar \nu ] \simeq 45  \% $,
Br[$Z^\prime \rightarrow \ell \bar \ell ] \simeq 13 \% $ and Br[$Z^\prime \rightarrow q \bar q ] \simeq 42  \% $. Thus, when we compare the leptonic $Z^{\prime}$ branching ratio with the SM Br[$Z\rightarrow \ell \bar \ell] \simeq 3 \% $, we can conclude that the search for the new gauge boson can be accessible via a clean dilepton signal at the LHC. Moreover, from the relation among the gauge boson masses, it is clear that our $Z^{\prime}$ can only decay into fermions and scalars, in contrast with the leptophobic versions where the channels involving the new ${\rm SU}(3)_L$ gauge bosons are present. Finally, by calculating the $Z^{\prime}$ width, we find that $\Gamma_{Z^{\prime}}$ is around  $5 \%  M_{Z^\prime}$.

Then, by considering the possibility of the ${\rm SU}(3)_L$ breaking scale being at the
${\cal O}(\unit{TeV})$, leading to a mass scale for 
the new gauge bosons of a few TeVs, we explore the production of a muon pair through
the decay of the heavy $Z^{\prime}$. It is clear that, from the experimental point of view, the new neutral gauge boson can be observed in
the invariant mass formed by the dilepton mass spectrum. The peak observed in the invariant mass distribution for the final particles,
over a smooth SM background, represents the evidence for new physics. Thus, in general, the experimental analysis searches for narrow
resonances where the experimental resolution is the dominant contribution to the observable width of a peak structure appearing over a
SM background. In this approach, theoretical cross section predictions for specific models are usually calculated in the
narrow width approximation.
Obviously, when the width is wide, the resonance appears as a
broad shape and can be almost flat around the $Z^{\prime}$ pole. 

Thus, within this narrow width approximation, we show below the invariant mass and transverse momentum $p_{T}$ distributions of
the emerging leptons in the processes $p + p \longrightarrow \mu^+ + \mu^- + X$ at 14 TeV, involving the  $Z^{\prime}$
of this new 3-3-1 version. We leave for a future work the study of the effects of a $Z^{\prime}$ with wide width, like the one predicted in the so-called minimal version.

To carry out our analysis, we consider the general Lagrangian for the neutral currents involving $Z$ and $Z^{\prime}$ contributions,
\begin{equation}
\mathcal{L}^{NC} =-\frac{g}{2 \cos\theta_W}\sum_{f} \Bigl[\bar
f\, \gamma^\mu\ (g_V + g_A \gamma^5)f \, Z_\mu+ \bar f \, \gamma^\mu\ (g^\prime_V + g^\prime_A \gamma^5)f \, { Z_\mu^\prime}\Bigr],
\end{equation}
where $f$ stands for leptons and quarks, $g$ is the weak coupling constant, and $g_V$,  $g_A$, $g^\prime_V$ and $g^\prime_A$, are the SM and 3-3-1 couplings which are presented in the
Table\;II, where we take the approximation $v/w\ll 1$ and assume no flavor mixing. Below the electroweak scale, the phenomenology predicted by the new model involving $\gamma$ and $Z$ coincides with the SM one. 
\begin{table}[h]
\label{table.2}
	\begin{footnotesize}
		\begin{center}
			\begin{tabular}{|c|c|c|c|c|}
				\hline 				
				&  $g_V$ & $g_A$   & $g^\prime_V$ & $g^\prime_A$
\\[.5ex]
				\hline
\rule{0cm}{2em}
				$Z \bar l l $ / $Z^{\prime} \bar l l $  &
				$\displaystyle{-\frac{1}{2} + 2\sin^2\theta_W}$ &
				$\displaystyle{-\frac{1}{2}}$~ &
				$\displaystyle{-\frac{1+2\sin^2\theta_W}{2\sqrt{3-4\sin^2\theta_W}}}$ &
				$\displaystyle{-\frac{1-2\sin^2\theta_W}{2\sqrt{3-4\sin^2\theta_W}}}$
\\[1em]			
				\hline
\rule{0cm}{2em}				
				$Z \bar u u$ / $Z^{\prime} \bar u u$ & 
				$\displaystyle{\frac{1}{2}-\frac{4\sin^2\theta_W}{3}}$& 
				$\displaystyle{\phantom{-}\frac{1}{2}}$~ &
				$\displaystyle{\frac{3+2\sin^2\theta_W}{{6\sqrt{3-4\sin^2\theta_W}}}}$  &
				$\displaystyle{\frac{1-2\sin^2\theta_W}{2\sqrt{3-4\sin^2\theta_W}}}$
\\[1em]							
				\hline
\rule{0cm}{2em}			
				$Z \bar d d$ / $Z^{\prime} \bar d d$&  
				$\displaystyle{-\frac{1}{2}+\frac{2\sin^2\theta_W}{3}}$  &  
				$\displaystyle{-\frac{1}{2}}$~ &
				$\displaystyle{\frac{\sqrt{3-4\sin^2\theta_W}}{6}}$   & 
				$\displaystyle{\frac{1}{6\sqrt{3-4\sin^2\theta_W}}}$ 	
\\[1em]
				\hline
			\end{tabular}
		\end{center}
	\end{footnotesize}
	\caption{The vector and axial couplings of $Z$ and $Z^{\prime}$ to leptons ($e$, $\mu$ and $\tau$) and quarks ($u$ and $d$) in the 3-3-1 models. $\theta_W$ is the Weinberg angle.}
\end{table}

By following previous studies on $Z^{\prime}$ concerning strategies for the identification of this particle 
on the muon channel \cite{EU, BELA, ELE}, as well as the last ATLAS report \cite{ATLAS}, 
we have applied some cuts in order to obtain clear distributions for the invariant masses and transverse momentum of the 
final muons. 
In agreement with the ATLAS detector performance, the cuts adopted for the pseudorapidity and the 
transverse momentum of the muons are: $\vert \eta\vert < 2.5$ and  $p_T > 30$ GeV. For the invariant mass of the
muon pair, we have used a strong cut ($M_{\mu\, \mu} > 1000$ GeV) in order to suppress the SM background.

In our simulations, we have made use of the CompHep~\cite{COM} and the MadAnalysis~\cite{MAD} packages and adopted the
CTEQ6L~\cite{CETQ} parton distribution functions set, evaluated at the $\sqrt{\hat{s}}$ factorization/re\-nor\-ma\-li\-za\-tion scale, 
{\it i.e.}, the center-of-mass energy at the parton level.

\begin{figure}[h]
	\center
	\resizebox{0.8\textwidth}{!}{\includegraphics[height=.72\textheight]{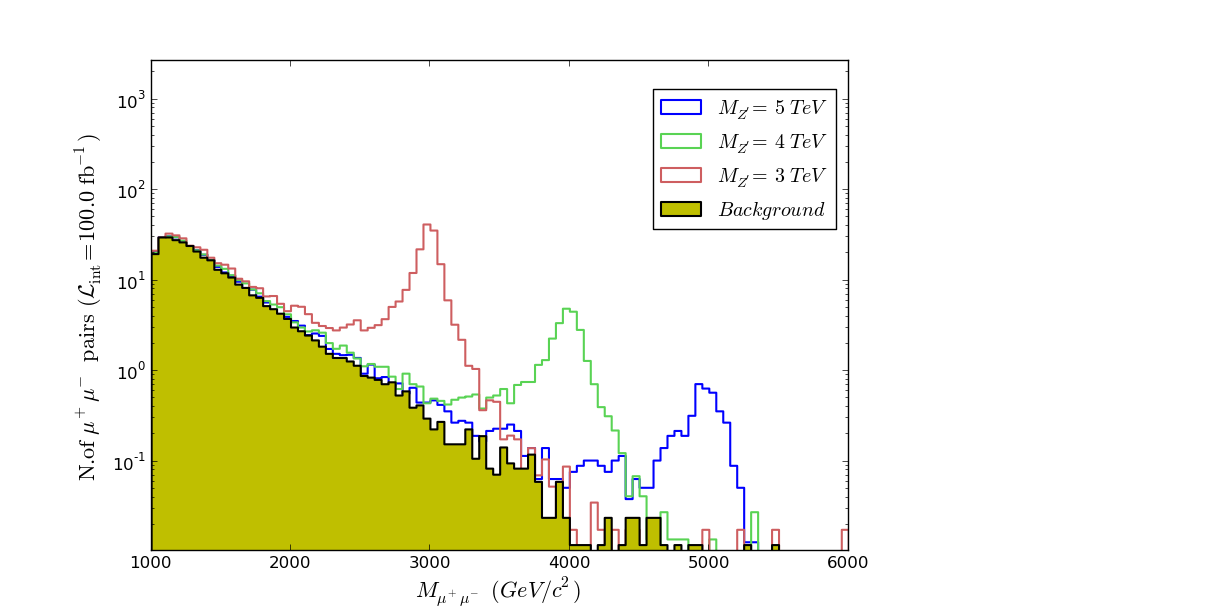}}
%
	\resizebox{0.8\textwidth}{!}{\includegraphics[height=.72\textheight]{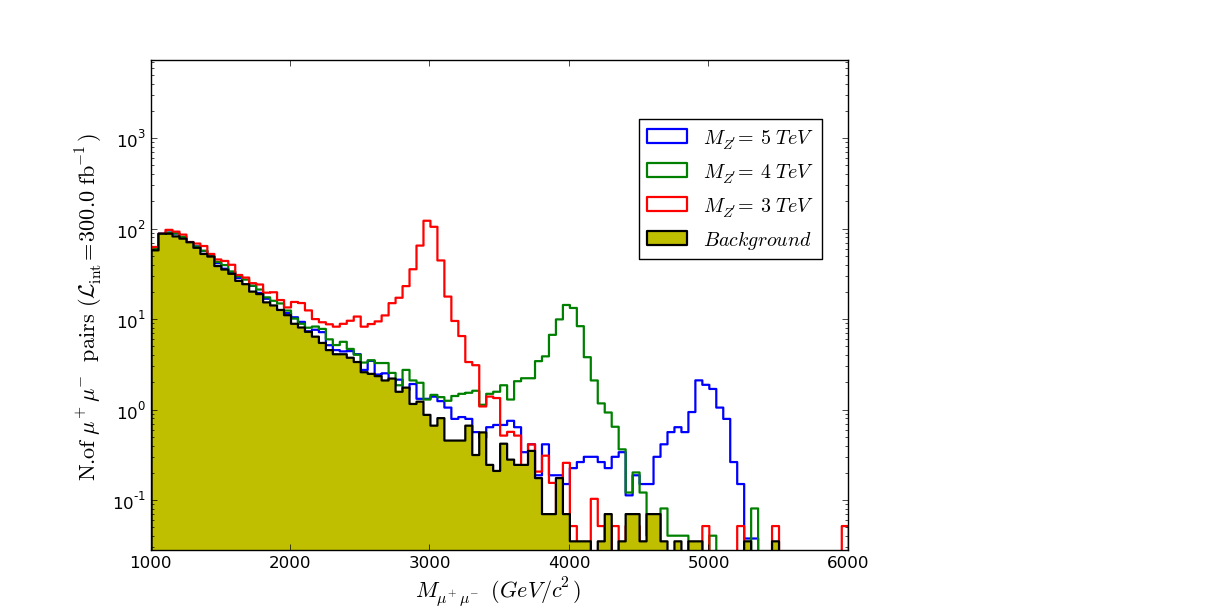}}
	\caption{Number of $\mu^{+} \mu^{-}$ pairs as a function of three representative $Z^\prime$ masses at the 14 TeV LHC with	${\cal L} = 100$ fb$^{-1}$ (upper panel) and ${\cal L} = 300$ fb$^{-1}$ (lower panel).}
	\label{invm}
\end{figure}

Upon assuming 3, 4 and 5 TeV for the $Z^{\prime}$ mass, we observe the resonance peaks around the respective masses 
in the invariant mass distribution. If we consider two values for the projected LHC integrated luminosity (${\cal L} = 100$ fb$^{-1}$ and 
${\cal L} = 300$ fb$^{-1}$) at $\sqrt{s}= 14$ TeV, we obtain the number of events as shown in Fig.\,\ref{invm}.  
As the width of the heavy boson satisfies the relation $\Gamma_{Z^{\prime}} \sim 5 \%  M_{Z^\prime}$, our results are in accordance
with Ref. \cite{ELE}. 

\begin{figure}
	\center
	\resizebox{0.8\textwidth}{!}{%
		\includegraphics[height=.72\textheight]{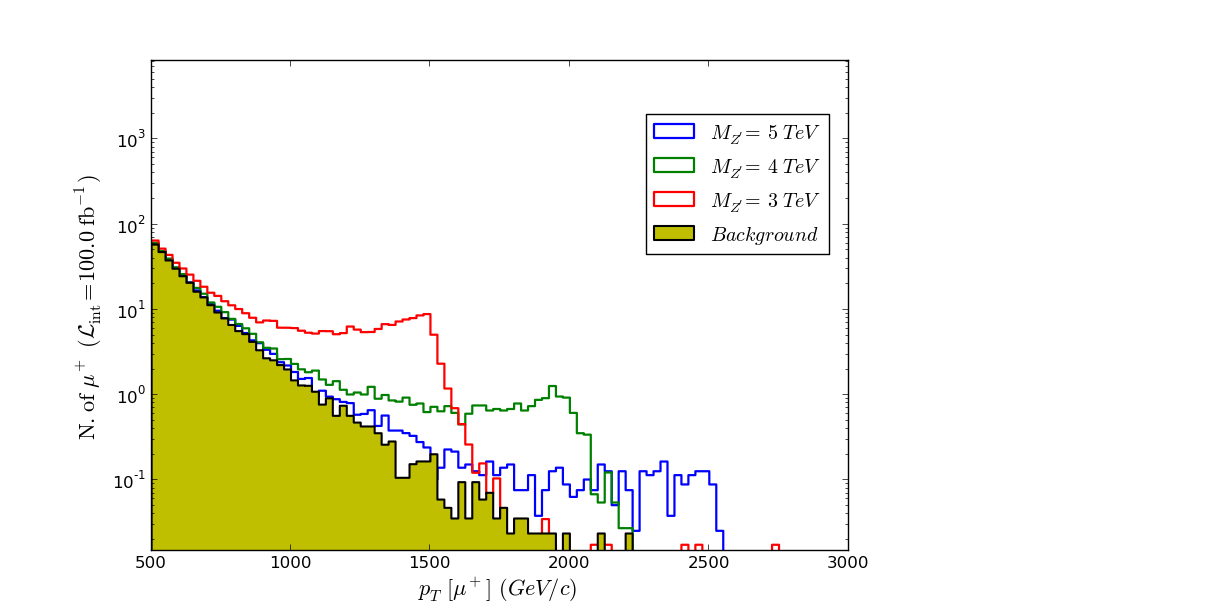}}
	\resizebox{0.8\textwidth}{!}{%
		\includegraphics[height=.72\textheight]{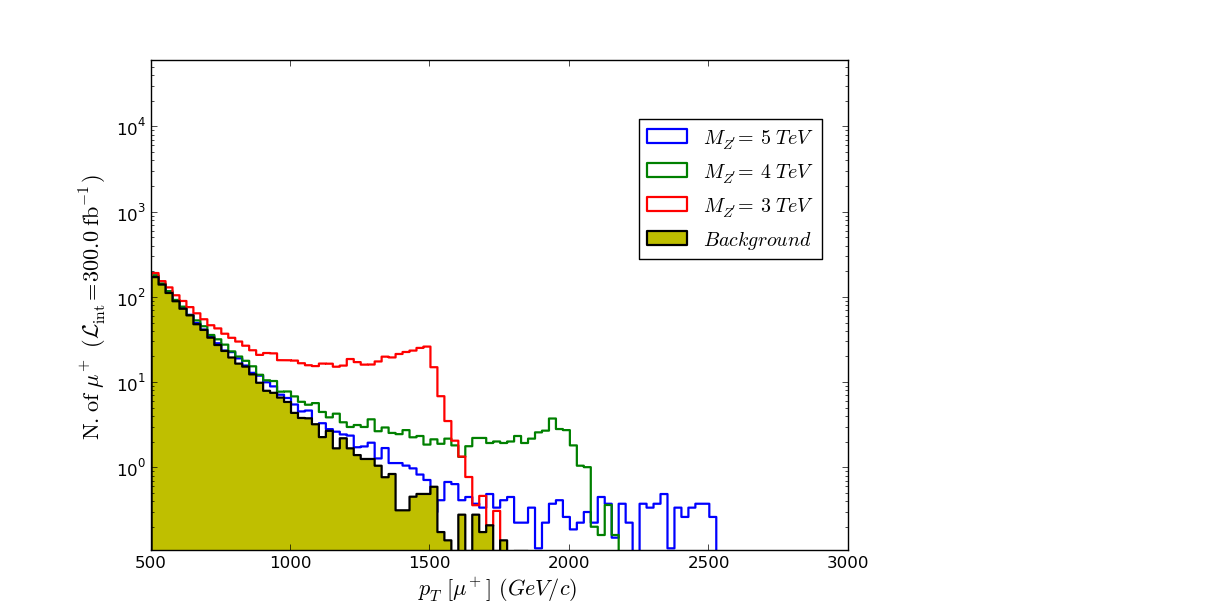}}
	\caption{Number of $\mu^{+}$ as a function of the $p_T$ of the emerging $\mu$ at the 14 TeV LHC with ${\cal L} = 100$ fb$^{-1}$ (upper panel) and ${\cal L} = 300$ fb$^{-1}$ (lower panel).
	}
	\label{invm:2}
\end{figure}

In order to identify the new gauge boson, we also consider the muon $p_T$ distribution, where two peaks are expected to appear,
corresponding to one half of the resonance masses ($M_{Z}$, $M_{Z^\prime}$) but, in this case, due to the invariant mass cut adopted,
the first peak moves justly to one half of this cut ($\sim 500 $GeV ). By applying an additional cut in the muon transverse momentum,
$p_T > 500$ GeV, we obtain the distributions shown in the Fig.\,\ref{invm:2}.
We can identify the peak around the $M_{Z^\prime}$/2 values,
with $M_{Z^\prime}$ = 3, 4 and 5 TeV, respectively. It is clear that, for $M_{Z^\prime}$ = 3 and 4 TeV, the peak is well defined, and for
greater masses, the peak is smooth like the SM background. 
Then, with a stronger cut on the invariant mass or on the muon $p_T$, we might obtain a clearer peak for higher masses in the $p_T$ distribution. 

Therefore, as claimed in \cite{EU,ELE}, we can use the transverse momentum distribution of the
final muons as an additional tool to distinguish the signal coming from the $Z^\prime$ from the SM background.
Thus, by a simple analysis of the invariant masses and transverse momentum of the the final $\mu$, we get clear signals for the $Z^{\prime}$. Moreover, by considering the projected luminosities for the LHC run-II, we obtain a considerable number
of events revealing the existence of the new gauge boson.
In a recent work \cite{CAO}, a phenomenological analysis of the  $Z^\prime$ has been performed for the version considered in this paper. The authors have suggested that the possibility of detecting the new particle can be achieved by considering just the production of a pair of leptons. From their analysis using, for example, the Forward-Backward asymmetry, they have concluded that at the LHC at 14 TeV, it is possible to identify the $Z^\prime$ boson.
Our strategy is a little different from the one used by those authors. We use the invariant mass distributions and the transverse
momentum of the final leptons, in order to distinguish the signal from the SM background.
In any case, our conclusions are similar regarding the possibility of discovering the $Z^\prime$ in the run II of the LHC.

It is beyond the scope of our work to make a detailed analysis of the final states, including $Z-Z^{\prime}$ interferences, detector efficiencies, hadronization, etc. However, based on our results, it is not hard to establish the existence or to exclude the 
$Z^\prime$ predicted by this model. Finally, a complete analysis involving the $Z^{\prime}$ predicted by different versions of the 3-3-1 model within the next stage of the LHC energy is mandatory, but we postpone this study to a future work.

\section{Conclusions}
\label{conclusions}

In this work we have presented a version of the 3-3-1 model, defined by $\beta=1/\sqrt 3$ in 
the electric charge operator in Eq.~\eqref{Qop}, which at low energies contains only 
two scalar triplets in order to achieve the correct breakdown of gauge symmetries. 
Eight out of the twelve degrees of freedom contained in the scalar triplets are absorbed 
in the longitudinal components of the vector bosons $Z$, $W^\pm$, $Z'$, $V^\pm$, $V^0$, $V^{0\dagger}$. This leaves three spin-0 bosons in the particle spectrum, 
two neutral CP even scalars $h$, $H$, and a charged scalar $\varphi^\pm$. 
The neutral scalar, $h$, is identified with the discovered Higgs boson, 
and gets its mass at the scale $v$, related to the ${\rm SU}(2)_L\otimes {\rm U}(1)_Y\rightarrow$
${\rm U}(1)_Q$ symmetry breakdown. Both the neutral, $H$, and the charged, $\varphi^\pm$, 
scalars are supposedly heavier, since they get their masses at the scale $w$, associated 
with the ${\rm SU}(3)_L\otimes{\rm U}(1)_X\rightarrow{\rm SU}(2)_L\otimes{\rm U}(1)_Y$ symmetry breakdown. 
In comparison to other Standard Model extensions, such as the two Higgs doublet
model for example, our construction has a smaller number of extra scalars
at the TeV scale because no CP odd neutral state is part of the spectrum.
For the same reason, FCNC mediated by scalars are very much suppressed
and absent in the limit of no mixing between SM fermions and heavy fermions.

The model has three extra charged leptons, ${\cal E}_i$, two up-type quarks ${\cal U}_a$, and 
one down-type quark ${\cal D}$, beyond the Standard Model fermion content. 
Although the model with just two scalar triplets has a 
consistent pattern of gauge symmetry breakdown to the electromagnetic factor ${\rm U}(1)_Q$, 
some of the standard fermionic fields remain massless due to a residual global 
${\rm U}(1)_{\cal G}$ symmetry which, as we observed in Eq.~(\ref{SG}), involves diagonal 
generators of spontaneously broken gauge symmetries plus a sort of Peccei-Quinn symmetry. This ${\rm U}(1)_{\cal G}$ symmetry seems to be a common feature of 3-3-1 models 
with just two scalar triplets and has also been identified in another version of the model~\cite{Montero:2014uya}. 
To overcome this problem, we have introduced a heavy scalar triplet with mass~$M\gg w$, 
which is integrated out from the low energy theory leaving it with effective operators 
breaking ${\rm U}(1)_{\cal G}$ explicitly, completing the mass generation mechanism for 
the fermions.
As we have shown, the effective operators furnish a less fine-tuned mass generation for leptons 
and up-type quarks compared to the Standard Model. Such a mechanism, however, does not 
work as naturally for the standard down-quark mass hierarchy
and a solution based on an additional $\ZZ_2$ symmetry has also been provided.
Natural hierarchies between the heavy quarks and the third family quarks, and between the third
family and the lighter two families, arise, and this feature naturally suppresses the mixing
between them leading to suppressed FCNC interactions.

From the phenomenological standpoint, we have explored the possibility to discover the predicted $Z^{\prime}$ by considering the leptonic decay channel within
the LHC energy regime. By making a simple analysis involving the invariant masses 
and transverse momentum of the final muons, and by selecting appropriate cuts for the final states,
we have concluded that clear signals can reveal the presence of the new neutral gauge boson. 
If we take the projected integrated luminosities for the next LHC phase, we find a considerable number of events for processes involving
$Z^{\prime}$ and the final muons, which could confirm one of the predictions of the model. Moreover, other potential tests involve the pair and single production 
of the new leptons and quarks, in addition to the $V^\pm$, $V^{0}$, $V^{0\dagger}$ vector bosons. For the minimal scalar 
sector, containing only an extra Higgs and a charged scalar, the production of H via gluon fusion, {\it i.e.} $g g \rightarrow H $, 
and the analysis of the final states $ b \bar b b \bar b $, $ b \bar b \tau \tau$  and $ b \bar b \gamma \gamma$ 
represent an excellent prospect for the discovery or exclusion of the new neutral scalar state. On the other hand,
the associated productions of the charged Higgs with a top quark and with a W boson,  via the partonic processes $b g \rightarrow t H^-$ 
and $ b \bar b \rightarrow  H^- W^+ $, can be also tested at the LHC.
Thus, considering both the theoretical and the phenomenological aspects presented, this new model is surely worth our attention in further studies.

\acknowledgments
This research was partially supported by the Conselho Nacional de Desenvolvimento Cient\'{\i}fico e Tecnol\'ogico (CNPq), by grants 306636/2016-6 (A.G.D.) and
308578/2016-3 (C.C.N.).
Financial support by Funda\c{c}\~{a}o de Amparo \`{a} Pesquisa do Estado de S\~ao Paulo (FAPESP) is also acknowledged under the grants 2013/22079-8 (A.G.D. and C.C.N.) and 2014/19164-6 (C.C.N.).
J.L. and R.O. would like to thank CAPES (Coordena\c{c}\~ao de Aperfei\c{c}oamento de Pessoal de N\'{\i}vel Superior), and W.C.V. thanks UFABC, for the financial support.

\appendix
\section{Higher order operators}
\label{ap:op}

We collect here the terms that we have omitted in \eqref{L:yukawa:2+1}:
\eqali{
-\lag_0&\supset
-\ \bar{q}_{3L}\phi_1 \ums[\sqrt{2}]{w}M_D D_R
+\bar{q}_{aL}\tphi_1\ums[\sqrt{2}]{w}M_{U_a}U_{aR}
\cr&~~
+\bar{D}_L\chi_3^0\ums[\sqrt{2}]{w}M_D D_R
+\bar{U}_{aL}\chi_3^{0*}\ums[\sqrt{2}]{w}M_{U_a} U_{aR}
\cr&~~
-\ \frac{\phi_2^\dag\phi_1}{\Lambda}\bar{U}_{aL}
\ums[\sqrt{2}]{\eps}\big[(M_u)_{ai}u_{iR}+(M_{uU})_{ab}U_{bR})\big]
\cr&~~
+\ \frac{\phi_1^\dag\phi_2}{\Lambda}\bar{D}_{L}
\ums[\sqrt{2}]{\eps}\big[m_b d_{3R} + (M_{dD})_3D_{R})\big]
+h.c.
}
\eqali{
-\lag_{\pm}&\supset \bar{D}_L\ums[\sqrt{2}]{v}\big[m_t u_{3R} + (M_{uU})_{3b}U_{bR}\big]
\rho_3^-
\cr&~~
+\ 
\bar{U}_{aL}\ums[\sqrt{2}]{\eps'}\big[(M_d)_{ai}d_{iR}+(M_{dD})_{a}D_{R})\big]\rho_3^+
+h.c.
}
Note that there are also effective Yukawa interactions involving $\rho_3^{\pm}$ in
\eqref{L:yukawa:2+1} coming from the terms within $\phi_3$.


\end{document}